\documentclass[pra,twocolumn,showpacs,preprintnumbers,amsmath,amssymb]{revtex4-1}
\usepackage{multirow}
\usepackage{graphicx}
\usepackage{dcolumn}
\usepackage{bm}
\usepackage{psfrag}
\usepackage{epsfig}
\usepackage{amsmath}
\usepackage{amssymb}
\usepackage{MnSymbol}
\usepackage{color}
\usepackage{bbm}
\usepackage[FIGTOPCAP,raggedright,nooneline]{subfigure}
\usepackage{verbatim}
 \usepackage[applemac]{inputenc}
\usepackage{textcmds} 
\usepackage{tikz}
\usepackage{natbib}

\newcommand{\ignore}[1]{}

\begin{document}

%
%
%
%

\title{Quantum acousto-optic control of light-matter interactions in nanophotonic networks}

\author{G. Calaj\'o$^{1,2}$,  M. J. A. Schuetz$^3$, H. Pichler$^{3,4}$, M. D. Lukin$^3$, P. Schneeweiss$^1$, J. Volz$^1$, P. Rabl$^1$}
\affiliation{$^1$ Vienna Center for Quantum Science and Technology,
Atominstitut, TU Wien, 1020 Vienna, Austria}
\affiliation{$^2$ ICFO-Institut de Ciencies Fotoniques, The Barcelona Institute of
Science and Technology, 08860 Castelldefels (Barcelona), Spain}
\affiliation{$^3$  Department of Physics, Harvard University, Cambridge, MA 02138, USA}
\affiliation{$^4$  ITAMP, Harvard-Smithsonian Center for Astrophysics, Cambridge, MA 02138, USA}

\date{\today}

\begin{abstract}

We analyze the coupling of atoms or atom-like emitters to nanophotonic waveguides in the presence of propagating acoustic waves. Specifically, we show that strong index modulations induced by such waves can drastically modify the effective photonic density of states and thereby influence the strength, the directionality, as well as the overall characteristics of photon emission and absorption processes. These effects enable a complete dynamical control of light-matter interactions in waveguide structures, which even in a two dimensional system can be used to efficiently exchange individual photons along selected directions and with a very high fidelity. Such a quantum acousto-optical control provides a versatile tool for various quantum networking applications ranging from the distribution of entanglement via directional emitter-emitter interactions to the routing of individual photonic quantum states via acoustic conveyor belts. 
\end{abstract}

\maketitle

%
%

\section{Introduction}

Optical signals can be transmitted through waveguides and fibers without being significantly degraded by the presence of acoustic excitations in the material. This property  can be attributed to the vast difference in frequency and propagation speed, which makes a direct coupling of photons and phonons very inefficient. Nevertheless, residual Brillouin scattering~\cite{Boyd,Bai2018}, where photons are scattered into other orthogonal modes by simultaneously emitting or absorbing phonons, still constitutes a major limitation for optical communication~\cite{Ippen1972,Slavik2010}, in particular when operating at higher power.  
In addition, has been suggested of using intense acoustic waves for the control of weak optical fields, 
 for example, for frequency conversion~\cite{Reed2003,Seddon2003}, on-chip phase modulation~\cite{Fan2016}, or non-reciprocal scattering of optical beams~\cite{Kang2011,Sohn2018}. In particular, in nanophotonic structures, where photons and phonons are both strongly confined~\cite{Dainese2006,Eichenfield2009,Merklein2015,Safavi-Naeni2018,DeLima2018}, such techniques represent a promising alternative to Kerr- or electro-optical modulation techniques for manipulating light~\cite{Yanik_FanPrl_1,Yanik_FanPrl_2,Notomi2006,Preble2007,Yu_Fan_nat,Lira2012,Calajo2017,Minkov2018}.

In this work we investigate the use of strong running acoustic waves for the control of nanophotonic quantum networks, where not only the propagation of single photons, but also their interaction with stationary emitters is of paramount importance. The basic idea is illustrated in Fig.~\ref{Fig1_Setup}(a), which shows a generic waveguide QED setting with multiple atoms, quantum dots or defect centers  that are strongly coupled to a one dimensional (1D) photonic channel~\cite{ReitzPRL2013,Hung2013,Thompson2013,Yalla2014,Arcari2014,review-lodahl,Goban2014,Hood2016,Pablo2017,Darrick_rev_mod,Lodahl_rev2}.
In engineered photonic crystal structures or near the edge of a propagation band, the group velocity of photons is considerably reduced, which can enhance and modify the coupling to guided modes~\cite{John1994,Kofman,lambro,Khurgin2010,Nori,Longo,Palma,Calajo2016,Tao}.
 However, in the resulting decay neither the shape nor the direction of the emitted photon is controlled, which prevents an efficient reabsorption of this photon by a second emitter. This pictures changes in the presence of strong index modulations induced by a propagating acoustic wave, which creates a moving lattice potential for the optical field. If sufficiently strong, this potential can confine and drag photons along, which induces a broad-band modification and, in particular, a left-right asymmetry in the effective photonic density of states, as experienced  by the emitter.  As a result, this method allows one to control both the rate as well as the direction of the emitted photons by simply adjusting the amplitude of the applied acoustic wave.

The ability to emit photons along a single direction and with a specified temporal shape is an essential requirement for the implementation of deterministic quantum communication protocols in scalable photonic networks~\cite{Cirac1997,Kimble2008}. Our analysis shows that acousto-optical control of emitter-photon interactions in waveguide QED systems provides a general method to achieve this tunability, without relying on specific level schemes or near-field effects~\cite{Luxmoore2013,Petersen2014,Mitsch2014,Sollner2015,ChiralRev}. Importantly, this technique can  be applied in two dimensional (2D) settings, where usually the dispersion of emitted photons into random directions prevents efficient interactions between emitters that are more than a few wavelengths apart. Acoustic control can be used to overcome this limitation and to implement long-range emitter-emitter interactions by channeling the emitted photons into a directed, strongly focused beam. Therefore, this method can be applied to extend concepts from non-reciprocal optics~\cite{Jalas2013} and chiral waveguide QED~\cite{ChiralRev} to higher dimensional scenarios, where quantum networks with a higher degree of connectivity as well as new types of quantum-optical and many-body phenomena can be explored.

This paper is structured as follows. In Sec.~\ref{sec model} we introduce a generic model for emitters coupled to an acousto-optical waveguide (AOW), which we use in Sec.~\ref{Sec. em_control} to derive a theory of spontaneous emission in the presence of strong acoustic waves. In Sec.~\ref{sec_QN} we then discuss several quantum networking applications based on acoustic control techniques in the weak- and strong-coupling regime. In Sec.~\ref{sec:2D} we extend our model to 2D waveguides and discuss the emergence of acoustically-induced directional emitter-emitter interactions in this setting. Finally, in Sec.~\ref{sec:Implementation} we discuss potential experimental settings for observing these effects with atoms or defect centers coupled to photonic crystal structures. 
In Sec.~\ref{sec conclusion} we summarize our results and discuss future directions of research.

\section{Model}\label{sec model}
\begin{figure}
  \centering
  \includegraphics[width=\columnwidth]{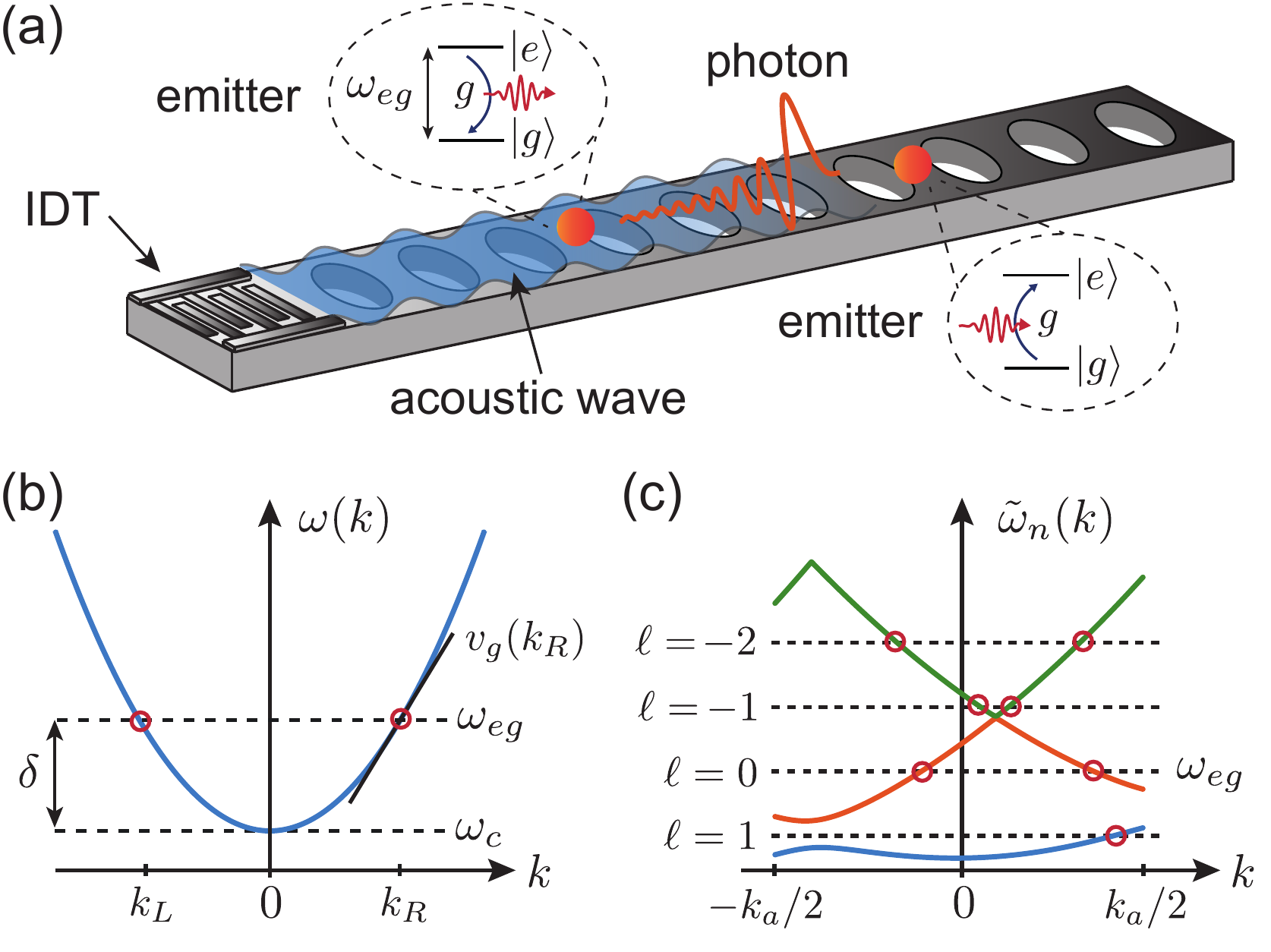}
   \caption{ \textit{Acousto-optical waveguide.}---(a) Sketch of a waveguide QED setup with multiple two-level emitters coupled to the field of a 1D photonic channel. A propagating acoustic wave, which is launched into the waveguide, for example, by an interdigital transducer (IDT), creates a strong modulation of the refractive index and modifies the photon emission and absorption properties. (b) Dispersion relation $\omega(k)$ of the unperturbed waveguide, which is assumed to be approximately quadratic above the cutoff frequency  $\omega_c$. In this case the emission rates $\Gamma_{R,L}\sim 1/|v_g(k_{R,L})|$ into right- and left-propagating modes are identical. (c) In the presence of the acoustic wave, the photon emission is instead determined by the deformed Floquet quasi-energy bands $\tilde \omega_n(k)$, which are  plotted for $V_a/E_r=0.2$ and $\Omega/\Omega_r=0.4$. For a finite speed of sound, $\tilde \omega_n(k)\neq \tilde \omega_n(-k)$ and photon emission becomes directional.  In (b) and (c) the red circles indicate the resonance conditions given in Eq.~\eqref{eq:ResonanceCondition}, which determine the set of wavevectors $k_\mu$ that contribute to the overall emission rate.   }
   \label{Fig1_Setup}
\end{figure}
We consider a generic setup as depicted in Fig.~\ref{Fig1_Setup}(a), where $N$ two-level atoms or solid-state emitters with ground state $|g\rangle$ and excited state $|e\rangle$ are coupled strongly to the  field of a 1D photonic waveguide. We assume that the emitters are dominantly coupled to photons of a single propagation band with a quadratic dispersion relation  $\omega(k)\simeq  \omega_c + \hbar k^2/(2m^*)$.  For conceptual simplicity we will primarily focus on homogeneous waveguides. In this case, $\omega_c$ is the cutoff frequency of a given transverse mode and $m^*\approx\omega_c \hbar n^2/c^2$, where $n$ is the refractive index, is the effective mass. However, as discussed in more detail in Sec.~\ref{sec:Implementation}, our analysis can be readily generalized to photonic crystal structures, where very strong couplings and much larger values of $m^*$,  i.e., a further reduction of the photonic group velocities, can be realized.

The waveguide is subject to a spatial and time-dependent modulation of the refractive index, $n(x,t)=n+\delta n(x,t)$, which creates an  effective potential  $V(x,t)\sim \delta n(x,t)$ for the photons~\cite{DeLima2005,DeLima2018,Sumetsky2011}. In this work we will specifically focus on strong index modulations induced by propagating acoustic waves via acousto-optical or optomechanical interactions~\cite{Fan2016,Kang2011,Sohn2018,DeLima2005,DeLima2018,Fuhrmann2011,Zoubi2016}, but our findings can be generalized to other electro-optical or Kerr-modulation schemes as well~\cite{Notomi2006,Preble2007,Lira2012}. 
The photons in the waveguide of total length $L\rightarrow \infty$ are then described by the Hamiltonian 
\begin{equation}\label{eq:Hf}
H_w(t)=\int_0^L dx \, \psi^\dag(x)\left(\hbar \omega_c-\frac{\hbar^2\partial^2}{2m^*\partial x^2}+V(x,t)\right)\psi(x),
\end{equation}
where $\psi(x)$ and $\psi^\dag(x)$ are bosonic field operators obeying $[\psi(x),\psi^\dag(x')]=\delta(x-x')$. The photons  interact with the emitters located at positions $x_i$ along the waveguide such that the Hamiltonian for the full system reads
\begin{equation}\label{eq:Htot}
\begin{split}
H=&\sum_{i=1}^{N}\hbar \omega_{eg}  |e\rangle_i\langle e|
+\hbar g\sum_{i=1}^{N}\left[\psi^\dag(x_i)\sigma_-^i +\sigma_+^i \psi(x_i)\right]+H_w(t).
\end{split}
\end{equation}
Here $\sigma_-=(\sigma_+)^\dag =|g\rangle\langle e|$ and it has been assumed that the transition frequency, $\omega_{eg}$, as well as the  coupling strength, $g(x_i)\simeq g$, is approximately the same for all emitters.

\section{Photon emission in acousto-optical waveguides}\label{Sec. em_control}
Let us first consider  the spontaneous emission of photons from a single emitter at position $x_1=0$, which is initially prepared in the excited state $|e\rangle$. Since the Hamiltonian~\eqref{eq:Htot} preserves the total number of excitations, the resulting system evolution is described by the wavefunction $|\Psi(t)\rangle = c_e(t) |e\rangle|{\rm vac}\rangle + \int dx\,  \phi(x,t) \psi^\dag (x)|g\rangle |{\rm vac}\rangle$. Here $p_e(t)=|c_e(t)|^2$ is the excited state probability and $\phi(x,t)$ is the amplitude of the emitted photonic wavepacket in position space. For $V=0$ and  $\omega_{eg}$  far from the cut-off frequency  we can use a conventional Wigner-Weisskopf approach to derive an effective equation for the decay of the excited state amplitude 
\begin{equation}\label{eq:dotce}
\dot c_e(t)= -i\omega_{eg} c_e(t) -\frac{\Gamma}{2}c_e(t),
\end{equation}
as well as for the right- and left-propagating emitted fields $\phi_{R/L}(t)=\lim_{\varepsilon\rightarrow 0^+}\phi(\pm \epsilon,t)$,   
\begin{equation}\label{eq:InOutFields}
\phi_{R/L}(t)= -i\sqrt{\Gamma_{R/L}/|v_g(k_{R/L})|} c_e(t).
\end{equation}
 In these equations, $\Gamma=\Gamma_L+\Gamma_R$ is the total decay rate and 
$\Gamma_{R}$ and $\Gamma_L$ are the rates of photons emitted to the right and to the left, respectively. For the unperturbed waveguide we recover the standard result, $\Gamma_R=\Gamma_L =g^2/|v_g(k_{R/L})|$. Here  $v_g(k)=\partial \omega(k)/\partial k$ is the group velocity and the two wavevectors $k_{R}=-k_{L}$ are  determined by the resonance condition $\omega_{eg}=\omega(k_{R/L})$.  

\subsection{Bloch-Floquet theory of spontaneous emission}\label{Sec Bloch-Floquet}

In the presence of the potential  $V(x,t)=V_a\cos[k_a(x-vt)]$ induced by a right-propagating acoustic wave with a speed of sound $v>0$, the photons experience an additional periodic modulation in space and time with frequency $\Omega=v k_a$ and wavelength $\lambda=2\pi/k_a$.  In this case it is convenient to change into the interaction picture with respect to the decoupled Hamiltonian $H_0=\hbar \omega_{eg}  |e\rangle\langle e|+H_w(t)$. In this new representation, the field operator $\psi_I(x,t)= U^\dag(t) \psi(x)U(t)$, where $U(t)=\mathcal{T}e^{-\frac{i}{\hbar}\int_0^t ds H_0(s)}$ and $\mathcal{T}$ is the time-ordering operator, can be written in terms of a Bloch-Floquet expansion as
\begin{equation}\label{expansion}
\psi_I(x,t)=\frac{1}{\sqrt{L}} \sum_{n,k}  e^{ikx} u_{nk}(x,t) a_{nk}.
\end{equation}
Here $k\in [-k_a/2,k_a/2)$ lies within the Brillouin zone (BZ) defined by the acoustic wavevector $k_a$ and the $a_{nk}$ $(a_{nk}^\dag)$ are bosonic annihilation (creation) operators. The  $u_{nk}(x+\lambda,t+2\pi/\Omega)=u_{nk}(x,t)$ are periodic functions, which  satisfy the differential equation
\begin{equation}\label{eq:unk_dot}
\dot u_{nk}=-\frac{i}{\hbar} \left[\hbar \omega_c -\frac{\hbar^2}{2m^*}\left(\frac{\partial}{\partial x}+ik\right)^2+V(x,t)\right]u_{nk},
\end{equation}
and can be decomposed as 
\begin{equation}\label{eq:unk_decomp}
u_{nk}(x,t)= e^{-i\tilde \omega_{n}(k) t}\sum_{\ell=-\infty}^{\infty} u_{nk}^{(\ell)} e^{i(k_ax-\Omega t)\ell}.
\end{equation} 
From the numerical solution of Eq.~\eqref{eq:unk_dot} we obtain a set of quasi-energy bands $\tilde \omega_n(k)$ [see Fig.~\ref{Fig1_Setup}(c)], which for a static potential just correspond to the usual Bloch bands. For $V_a/E_r\gtrsim1$, where $E_r=\hbar \Omega_r=\hbar^2k_a^2/(2m^*)$ is the photonic recoil energy, the lowest bands become well separated and their width decreases. 
Importantly, for finite propagation velocity $v$, we observe an asymmetric distortion of the quasi-energy bands, i.e.,  $\tilde \omega_n(k)\neq \tilde \omega_n(-k)$. 
 Therefore, the density of right- and left-propagating photonic states in an acousto-optical waveguide is no longer the same, which can give rise to directional emission of photons, as discussed below.

By using the decomposition~\eqref{eq:unk_decomp}, the remaining interaction Hamiltonian can be written as
\begin{equation}\label{H_eff}
\begin{split}
H_{I}(t) =\frac{\hbar g}{\sqrt{L}}\sum_{kn\ell} u_{nk}^{(\ell)}  e^{-i(\tilde \omega_n(k)+\Omega \ell-\omega_{eg})t} a_{nk} \sigma_+ + {\rm H.c.}  
\end{split}
\end{equation}
This expression shows that resonant interactions between the emitter and the field can occur at multiple wavevectors $k_\mu$,  which satisfy the resonance condition 
\begin{equation}\label{eq:ResonanceCondition}
\omega_{eg}=\tilde \omega_{n_\mu}(k_\mu)+\Omega \ell_\mu,
\end{equation}
for a band index $n_\mu$ and a Floquet index $\ell_\mu$ [see Fig.~\ref{Fig1_Setup}(c)]. The emission rate into modes around $k_\mu$ will depend on the coupling $\bar g_{\mu}= g u_{n_\mu k_\mu}^{(\ell_\mu)}$ and the quasi group velocity 
$\tilde v_{g,\mu}=\partial \tilde \omega_{n_\mu}(k)/\partial k|_{k=k_\mu}$. By summing over all resonant $k$-vectors we obtain the total emission rate $\Gamma=\Gamma_R+\Gamma_L$. The corresponding rates for emitting into right- and left-propagating modes are now given by 
[see App.~\ref{AppA}],
\begin{equation}\label{decay rate}
\Gamma_{R,L}=\sum_{\mu}\frac{ |\bar g_{\mu}|^2} {| \tilde v_{g,\mu}|}\theta[\pm \tilde v_{g,\mu}],
\end{equation}
where $\theta(x)$ denotes the Heaviside step function. 
 In the following
we introduce the characteristic decay rate $\Gamma_0=g^2 /|v_g(k_a/4)| =4\pi g_0^2 /\Omega_r$, where $g_0=g/\sqrt{\lambda}$ is the coupling strength between an emitter and a single photon of extent $\lambda$.  This rate corresponds to the rate of emission into the unperturbed waveguide at a frequency $\omega_{eg}=\omega(k_a/4)$ in the middle of the first BZ.

\subsection{Acoustically-induced directionality}\label{SecAIb}

\begin{figure}
  \centering
  \includegraphics[width=\columnwidth]{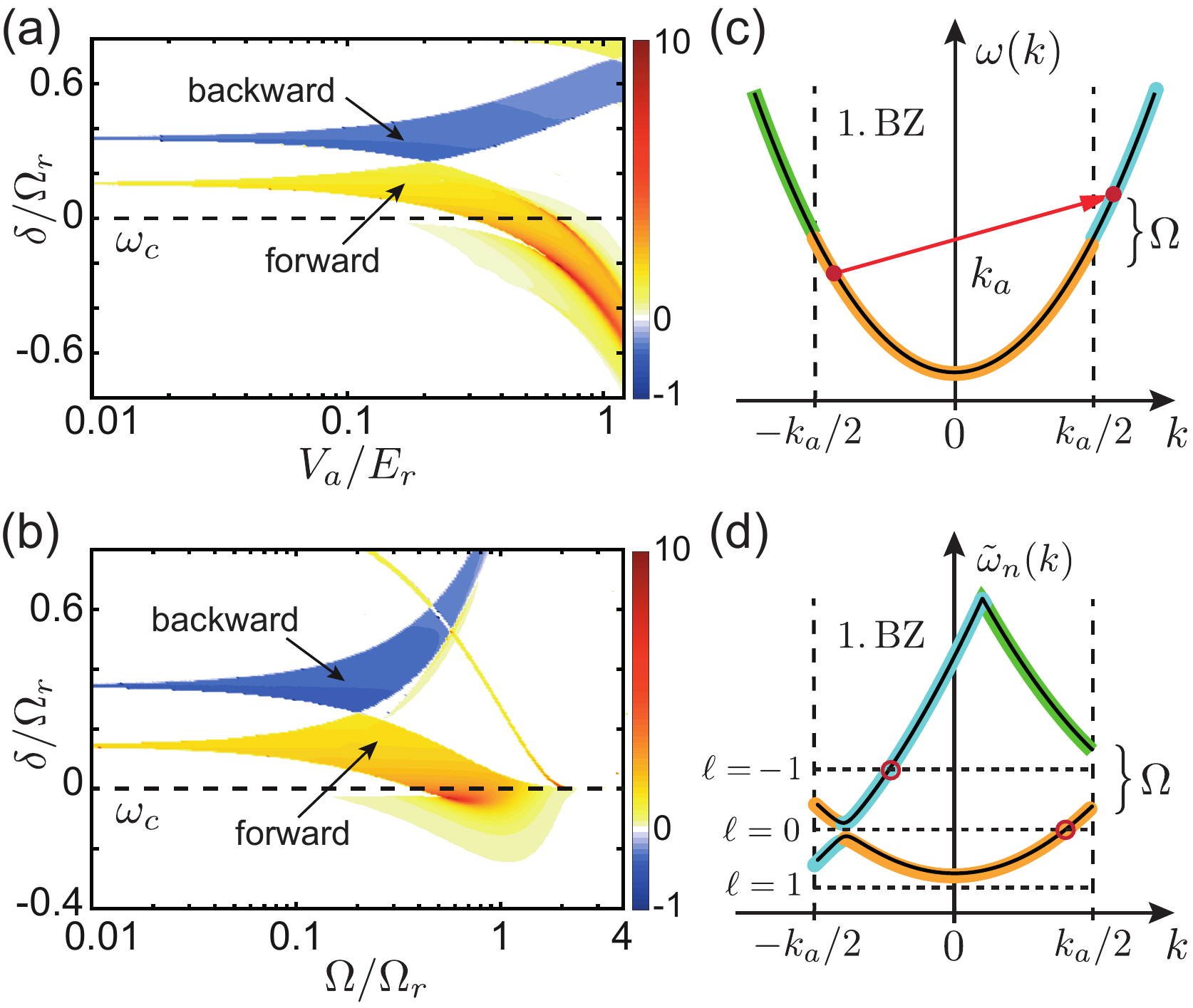}
   \caption{\textit{Directional photon emission.}---The directionality parameter $D=(\Gamma_R-\Gamma_L)/\Gamma_0$ is plotted as a function of the detuning of the emitter from the band edge,  $\delta=\omega_{eg}-\omega_c$, and in (a)  for varying $V_a$ and fixed $\Omega/\Omega_r=0.2$ and in (b) for varying $\Omega$ and fixed $V_a/E_r=0.2$. The  dashed black line indicates the position of the band edge of the unperturbed waveguide. Note that in both plots we have restricted the maximum value of the directionality parameter to $|D|\leq 10$ in order to avoid unphysical divergencies near the band edges, where the assumption of an exponential decay breaks down.  In addition, for frequencies outside the modified photonic band,  where the Wigner-Weisskopf approach is no longer valid, $D$ has been set to zero.
   (c) Illustration of a Brillouin scattering process between two modes $k$ and $k^\prime=k+k_a$ for a quadratic dispersion relation. For finite $V_a$ the coupling of modes in the vicinity of $k$ and $k^\prime$ leads to an avoided crossing in the quasi-energy bandstructure, as shown in (d). Here the main resonances [see Eq.~\eqref{eq:ResonanceCondition}] are indicated by red circles.
 }
  \label{Fig2_Directionality}
\end{figure}

As compared to the standard setting, the travelling acoustic wave imposes a preferred direction, thereby breaking the symmetry of
the band structure $\tilde \omega_n(k)$. As a result the emission rates into right- and left-propagating photons, as defined in Eq.~\eqref{decay rate}, will in general be different. This difference can be  quantified in terms of the directionality parameter $D=(\Gamma_R-\Gamma_L)/\Gamma_0$, which is plotted  in Fig.~\ref{Fig2_Directionality} (a) and (b) for various potential parameters. We see that for a weak acoustic perturbation and low frequencies, $V_a<\hbar\Omega\ll E_r$, an asymmetric emission occurs only at two specific resonances
\begin{equation}
\delta=\omega_{eg}-\omega_c \simeq \frac{\Omega_r}{4}\mp \frac{\Omega}{2},
\end{equation}
where $D$ can be both positive (the photons are emitted into the direction of the acoustic wave) or negative (the photons are emitted into the opposite direction). As indicated in Fig.~\ref{Fig2_Directionality}(c), these resonances arise from a Brillouin scattering process between modes $k$ and $k^\prime=k+k_a$ of the unperturbed waveguide. This scattering process is resonant only for wavevectors that satisfy $\omega(k)+\Omega=\omega(k^\prime)$ [indicated by the red arrow in Fig.~\ref{Fig2_Directionality}(c)], which for finite $V_a$ leads to an avoided crossing in the vicinity of these modes. When either $k_{R}$ or $k_L$ lies within this avoided crossing, the corresponding left- or right-propagating emission channel is suppressed and the emission becomes directional.

This mechanism can also be understood from the opening of a gap in the quasi-energy band-structure, which is defined within the first BZ associated with the acoustic wavevector $k_a$. Here one has to keep in mind that the acoustic modulation $\sim V_a e^{\pm i(k_ax-\Omega t)}$ induces Umklapp processes that connect neighboring Brillouin zones in Floquet sectors that differ by $\ell=\pm 1$. Therefore, as illustrated in Fig.~\ref{Fig2_Directionality}(d),  when folding the original dispersion relation into the first BZ, the individual branches must be simultaneously shifted in frequency by $\pm \Omega, \pm 2\Omega, ... $, in order to obtain the correct avoided crossings. This construction explains the resulting asymmetry of the quasi-energy bands, which is retained when the potential strength is increased [see Fig.~\ref{Fig1_Setup}(c)].  Note that in this quasi-energy picture higher $\ell$-resonances must be taken into account already for small values of $V_a$. For example, for the backward emission process (which takes place in the second BZ) the resonances at $\ell=\pm1$ are more important than the $\ell=0$ contribution.

\subsection{Photon-dragging regime} 
As evident from Fig.~\ref{Fig2_Directionality}(a) and (b), this simple Brilluoin-scattering picture no longer applies for stronger potentials, $V_a/E_r\gtrsim 0.1$. In this regime, the acoustic index modulation is already sufficiently strong to spatially confine the photons, meaning that the emitted photons are dragged along by the moving lattice potential, rather than just being reflected from it. As a consequence, the effective photonic density of states of the waveguide changes substantially over a wide range of optical frequencies and instead of individual resonances, broad windows of directional emission appear. Importantly, strong forward emission can now occur even for frequencies $\omega_{eg}<\omega_c$, where in the absence of the acoustic wave emission into waveguide modes is completely inhibited. These features vanish again for $\Omega \gtrsim \Omega_r$, where the modulation is already too fast to significantly influence the decay process.

The photon-dragging effect not only affects the directionality, but  can also significantly enhance the overall emission rate due to a strong reduction of the quasi group velocity $\tilde v_g$. This enhancement becomes most pronounced, when the reduced photonic group velocity of the static lattice matches the speed of sound. In this  case the  photons reside in the vicinity of the emitter for a very long time and therefore interact more efficiently.  
This effect is closely related to the appearance of non-perturbative features in the emission of Cherenkov photons into slow-light waveguides~\cite{Calajo2017MA}, where a similar enhancement of the coupling between co-propagating photons and atoms can occur.  Note, however, that the process of photons being emitted from a moving emitter and the emission of photons into a moving photonic lattice are in general not the same, since the presences of a periodic structure breaks Galilean invariance~\cite{Longhi2018}.

\subsection{Multi-emitter waveguide QED}\label{sec:ME}

The Bloch-Floquet theory for spontaneous emission discussed above can be readily generalized to settings where multiple emitters are placed along the waveguide. To account as well for additional external driving fields, such scenario can be modeled by an effective  equation of motion  for the reduced density operator of the emitters, $\rho$, 
as obtained after eliminating all the photonic modes within a standard Born-Markov approximation. Note that this approximation is valid away from any divergencies in the density of states associated with the quasi-energy bands, which no longer coincide with the divergency at the original band edge at $\delta=0$.
As a result of this derivation detailed in App.~\ref{AppB}, we obtain a master equation of the general form 
\begin{equation}\label{eq:MasterEq}
\dot \rho=-\frac{i}{\hbar}\left[ H_{e}, \rho\right]  + \sum_{i,j=1}^N \left[ A_{ij}  \left(\sigma_-^i\rho \sigma_+^j- \sigma_+^j\sigma_-^i\rho\right) +\rm H.c. \right].
\end{equation}
Here, the first term describes the coherent evolution of the individual, laser-driven emitters. In a frame rotating with the laser frequency $\omega_L$, it reads
\begin{equation}
H_e = \sum_{i=1}^N - \hbar\delta_L |e\rangle_i\langle e|   + \frac{\hbar\Omega_L}{2} \left(e^{i\varphi_i} \sigma_-^i + e^{-i\varphi_i}\sigma_+^i \right),
   \end{equation}
   where $\delta_L=\omega_L-\omega_{eg}$, $\Omega_L$ is the Rabi-frequency and the $\varphi_i$ are locally adjustable laser phases. The second term in Eq.~\eqref{eq:MasterEq} accounts for all decay and waveguide-mediated interaction processes, $\sim A_{ij} \sigma_+^j \sigma_-^i$, which arise  from the emission and reabsorption of photons between different emitters $i$ and $j$. The corresponding amplitudes are given by 
\begin{equation}\label{eq:Aij}
A_{ij}= \frac{\Gamma_{\rm ng}}{2}\delta_{ij}+ \sum_{\mu }\frac{|\bar g_{\mu}|^2 e^{ i(k_\mu+k_a\ell_{\mu})r_{ij}}}{|\tilde v_g(k_\mu)|}\theta\left[\tilde v_g(k_{\mu}) r_{ij} \right],
\end{equation}
where $r_{ij}=x_j-x_i$ and the index $\mu$ runs again over all resonant wavevectors, $k_\mu$. 

In Eq.~\eqref{eq:Aij}, the diagonal terms, $A_{ii}= (\Gamma+\Gamma_{\rm ng})/2$, describe the decay of each individual emitter, where we have included an additional rate $\Gamma_{\rm ng}$ to account for all other decay processes into non-guided modes. The general expression for $A_{ij}$ explicitly shows that not only the decay of each individual emitter, but also their mutual interactions can be strongly influenced by the applied acoustic modulation. This allows one, for example, to switch between a regular ($|A_{ij}|=|A_{ji}|$) and a fully chiral ($|A_{ji}|\ll |A_{ij}|$ for $x_j>x_i$) waveguide QED system in a dynamical and fully tunable  way, by simply varying the amplitude of the acoustic wave.

\section{Quantum networking  applications}\label{sec_QN}
In the previous section we have shown that the presence of strong acoustic waves can influence both the strength and the directionality of photon emission, or even open up a decay channel at frequencies where otherwise emission into guided modes would not be possible. The key feature is that such modification can be induced dynamically by simply changing the amplitude or direction of the acoustic modulation.  This level of control becomes an essential ingredient for various quantum communication schemes, where propagating photons are used to distribute quantum states or generate entanglement between multiple emitters along the waveguide~\cite{Cirac1997,Kimble2008,ChiralRev,Stannigel2012,Pichler2015,Gonzalez-Ballestero2015}. In this context it is not only important to control the emission of photons, but also to efficiently reabsorb these photons at a distant site. In the following we will illustrate some of the  possibilities that are offered by acoustic control schemes for optical quantum networking applications.

\subsection{Dynamical control of photon emission}

\begin{figure}
  \centering
  \includegraphics[width=\columnwidth]{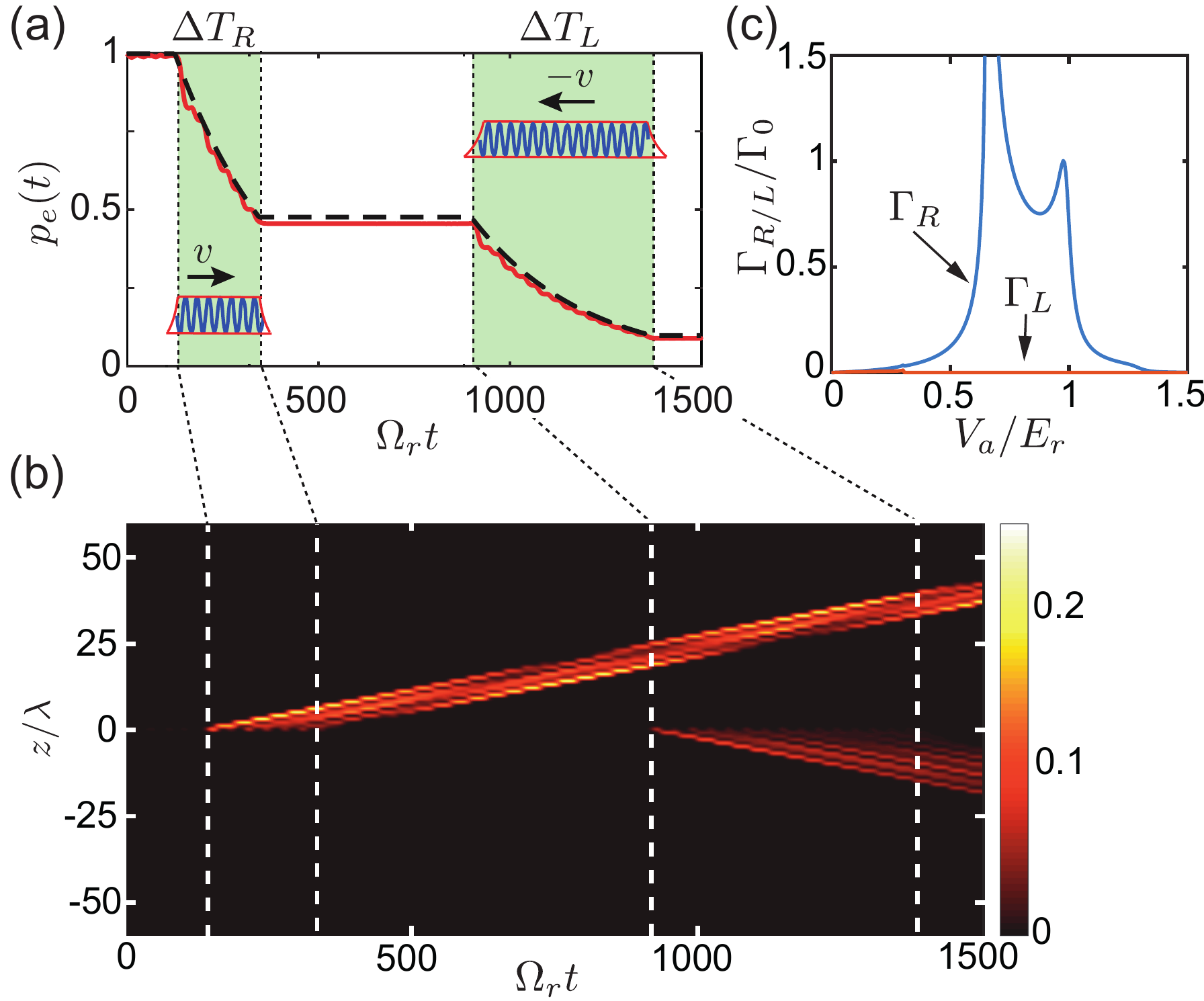}
   \caption{\textit{Acoustic emission control.}---(a) Plot of the excited state probability $p_e(t)$ for an emitter with a frequency $\omega_{eg}$ slightly below the band edge. During the time interval $\Delta T_R$ a right-propagating acoustic wave passes the emitter and induces a rapid decay. After  the acoustic wave has passed the emission is again inhibited until a second left-propagating acoustic wave packet induces another decay during the time interval $\Delta T_L$. The solid red line shows the results from a numerical simulation of the full Hamiltonian and the dashed line the results obtained from a Markovian theory with time-dependent decay rates $\Gamma_{R,L}(t)$ evaluated from Eq.~\eqref{decay rate}.  (b) Plot of the emitted photon wave packet $|\phi(x,t)|^2$, which shows that during the two time intervals the photon is emitted into different directions. The parameters for both plots are $V_a/E_r=0.8$,  $\Omega/\Omega_r=0.2$, $\delta/\Omega_r=-0.2$ and $g_0/\Omega_r=0.015$. (c) Dependence of  $\Gamma_{R}$ and $\Gamma_{L}$ on the potential strength $V_a$ for $\Omega/\Omega_r=0.2$ and $\delta/\Omega_r= -0.2$. Within the slowly-varying envelop approximation, this dependence can be use to achieve a time-dependent control of the emission rate $\Gamma_{R}(t)\gg \Gamma_{L}(t)$.}
  \label{Fig3_EmissionControl}
\end{figure}

As a first example, we illustrate in Fig.~\ref{Fig3_EmissionControl} the ability to dynamically control the emission properties of a single emitter by acoustic wave packets with varying propagation directions and amplitudes $V_a(t)$. 
Here we consider an emitter with a frequency well within the band gap, $\omega_{eg}<\omega_c$, such that initially emission into the unperturbed waveguide is strongly suppressed. During the time interval $\Delta T_R$ a right-propagating acoustic pulse with amplitude $V_a/E_r=0.8$ passes the emitter. According to Fig.~\ref{Fig2_Directionality}, under these conditions one expects a strong decay into right-propagating photons, which is clearly evident from the decay of $p_e(t)$ and the emitted photon wave packet shown in Fig.~\ref{Fig3_EmissionControl}(a) and (b), respectively. Once the acoustic wave packet has passed, the decay process stops half way in between. After a certain waiting time a second wave propagating in the opposite direction leads to a decay of the remaining population by emitting a photon to the left. Note that in the absence of other decay channels, the whole process is fully coherent and produces a superposition between a right- and a left-propagating photon.  

In Fig.~\ref{Fig3_EmissionControl}(a) the evolution of $p_e(t)$ is calculated from Eq.~\eqref{eq:dotce} with time-dependent rates $\Gamma_{R,L}(t)$. These rates are derived in a quasi-static approximation from the slowly-varying envelop of the modulation, $V_a(t)$, as depicted in Fig.~\ref{Fig3_EmissionControl}(c). This approximate theory is compared with an exact simulation of the emission process based on the full Hamiltonian~\eqref{eq:Htot}. We see that within the regime of validity, $g_0\ll \Omega_r$, the system dynamics is captured very well by the Markovian model. Therefore, this comparison shows that by slowly modulating the envelope of the acoustic wave, $V_a(t)$, a complete dynamical control over the emission rate combined with a high degree of directionality, $\Gamma_R(t)\gg \Gamma_L$, can be achieved. This  feature enables the emission of  photonic wave packets of arbitrary shape, or conversely, the absorption of arbitrarily shaped photons with close to unit efficiency. These are the central requirements for implementing deterministic quantum state transfer protocols between two separated emitters~\cite{Cirac1997}.

\subsection{Generation of stationary entangled states}\label{Sec entang}
To avoid precise pulse control, quantum correlations between multiple emitters can also be established under continuous driving conditions, as described by master equation~\eqref{eq:MasterEq}. In this case, the interplay between laser excitations and correlated decay processes into the waveguide can result in a non-trivial steady state, $\rho_0=\rho(t\rightarrow \infty)$, with a high degree of entanglement~\cite{ChiralRev,Stannigel2012,Pichler2015,Gonzalez-Ballestero2015}.  However, since the maximal amount of entanglement that can be reached by this approach depends crucially on the waveguide properties [characterized by the set of $A_{ij}$ in Eq.~\eqref{eq:Aij}], such schemes are not applicable in most conventional settings. By manipulating the $A_{ij}$ via strong acoustic waves, it is possible to overcome this limitation and to turn even a regular waveguide into an entanglement-mediating quantum channel.

\begin{figure}
  \centering
  \includegraphics[width=\columnwidth]{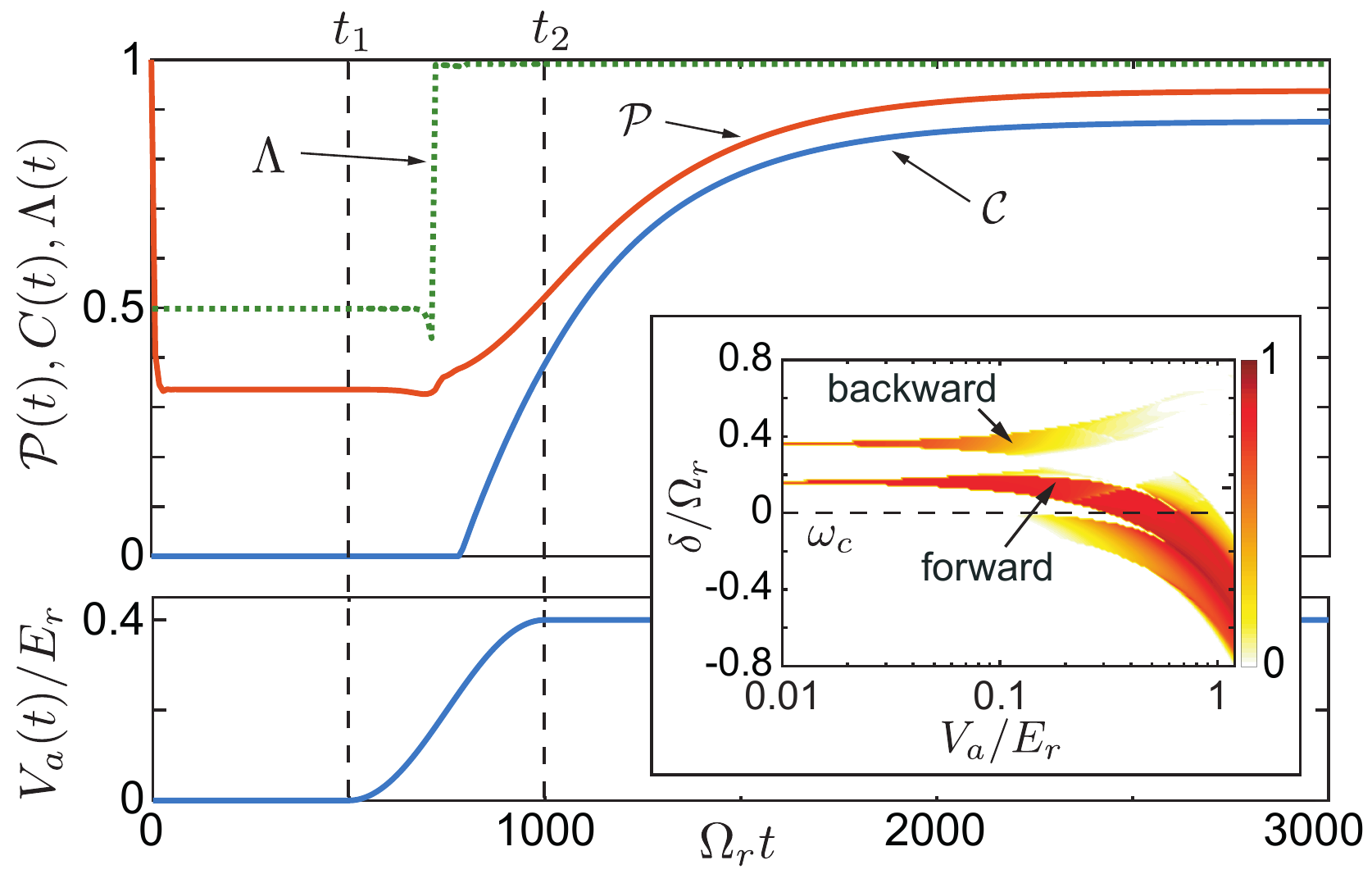}
   \caption{\textit{Steady-state entanglement.}---Evolution of the purity $\mathcal{P}$ and the concurrence $\mathcal{C}$ of the reduced density operator of two driven emitters coupled to a waveguide.   During times $t_1$ and $t_2$ the strength of the right-propagating acoustic potential, $V_a$, is gradually turned on and kept at a fixed value of $V_a/E_r=0.4$ afterwards (see lower plot). This tunes the waveguide into a regime, where the directional correlation parameter $\Lambda=|A_{12}|/(\Gamma+\Gamma_{\rm ng})$ (dotted line) is close to one. As a result, the system evolves into an almost pure steady state with a high degree of entanglement. For this plot we have assumed $\delta/\Omega_r=0.08$, $\Omega/\Omega_r=0.2$,   $g_0/\Omega_r=0.08$,  $\Gamma_{\rm ng}/\Gamma_0=0.001$, and we have set $\phi_i=- ik_R x_i$ to compensate for propagation phases.
 The inset shows the steady-state concurrence $\mathcal{C}(t\rightarrow \infty)$ as function of $\delta$ and $V_a$ for $\Omega/\Omega_r=0.2$ and $\Gamma_{\rm ng}/\Gamma_0=0.002$. For the simulations shown in the inset,  we fixed the Rabi frequency to a value of $\Omega_L=1.3\Gamma$ and averaged the resulting concurrence over different emitter separations $d=x_2-x_1$, to eliminate position-dependent interference effects.}
 \label{Fig4:Entanglement}
\end{figure}

To illustrate this concept, we consider in Fig.~\ref{Fig4:Entanglement} the case of two resonantly driven emitters with frequencies $\omega_{eg}>\omega_c$ within the propagation band of the unperturbed waveguide.  Although in this case the emitters can mutually exchange photons through the waveguide, many of these photons will simply be lost and the resulting steady state is  highly mixed and completely disentangled. As the acoustic modulation is gradually turned on, a directional emitter-emitter coupling, $|A_{12}|\gg |A_{21}|$, is established~\cite{Comment1}. After this point the system relaxes into an almost pure state, with a high degree of entanglement, expressed in terms of the concurrence $\mathcal{C}$~\cite{Wootters}. Indeed, in the limit of an ideal unidirectional quantum channel, where 
$A_{12}=\Gamma_R e^{ik_R (x_2-x_1)}$ and $A_{21}=0$, master equation~\eqref{eq:MasterEq} has a unique pure steady state, $\rho_0=|\psi_0\rangle\langle \psi_0|$~\cite{Stannigel2012,Pichler2015}, where
\begin{equation}\label{steady}
|\psi_0\rangle = \sqrt{\frac{\Gamma_R^2}{\Gamma_R^2+2\Omega_L^2}} \left( |gg\rangle  -i\frac{\sqrt{2}\Omega_L}{\Gamma_R} |S\rangle \right),
\end{equation}
is a superposition between the ground and the maximally entangled singlet state, $|S\rangle= (|ge\rangle-|eg\rangle)/\sqrt{2}$. 
As shown in Fig.~\ref{Fig4:Entanglement}, a steady state with a similar degree of entanglement can be achieved, even for a bi-directional, but modulated waveguide. This scheme for generating entangled steady states works  for a large range of parameters and both within the forward- and the backward-emission window (see inset). Note that by choosing a detuning $\delta<0$, the acoustic modulation can also be switched off after the steady state is reached, leaving behind a protected entangled state between two emitters inside the band gap~\cite{Bellomo}.

\subsection{An acoustic conveyor belt for light}\label{sec_mc}

\begin{figure}
  \centering
  \includegraphics[width=\columnwidth]{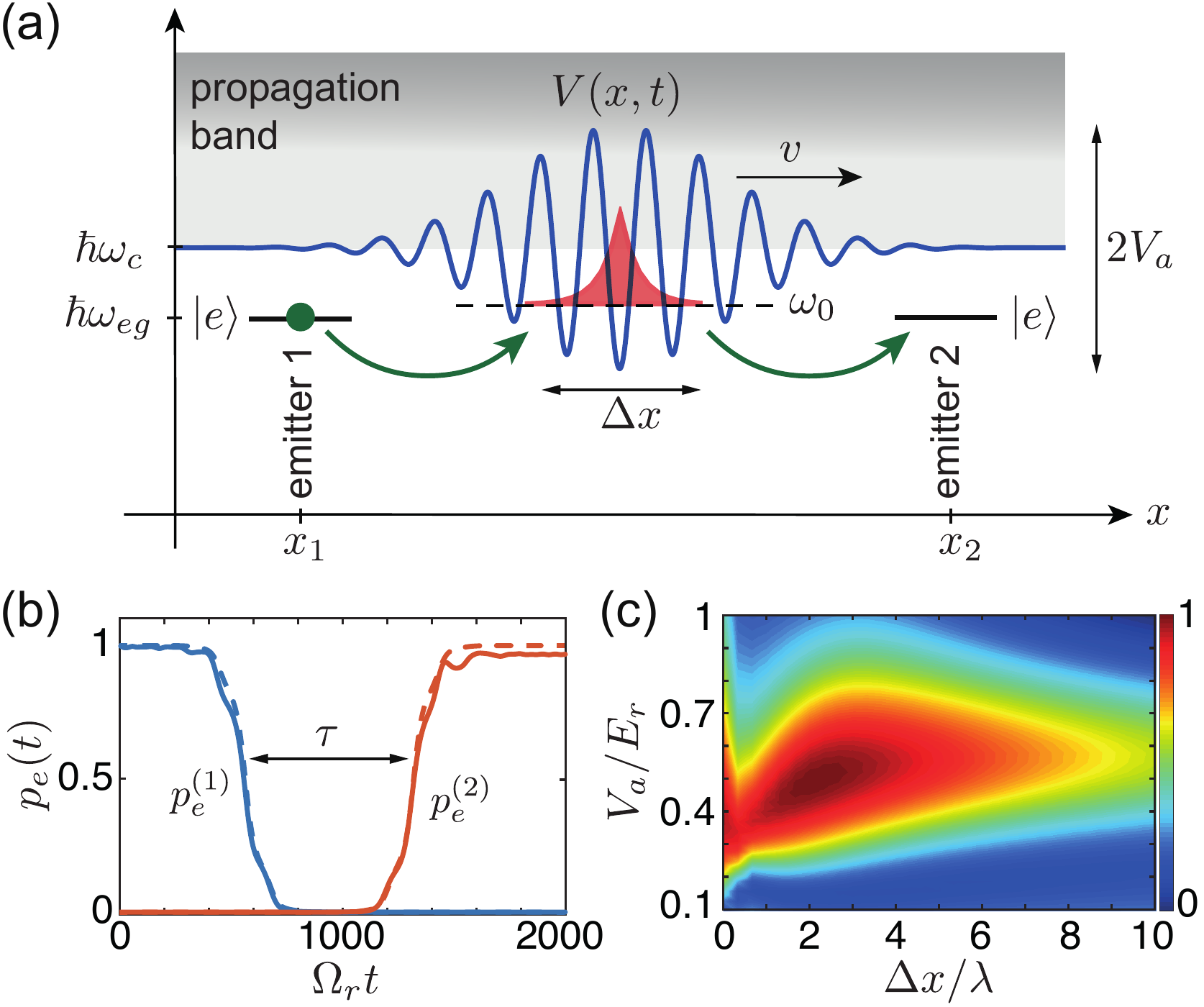}
   \caption{\textit{An acoustic conveyor belt.}---(a) Energy level diagram for the waveguide QED system in the presence of a propagating acoustic potential well $V(x,t)$, as defined in Eq.~\eqref{localized_pot}.  The excitation of the first emitter can be coherently converted into a single bound photon, which is dragged along by the acoustic wave and can successively interact with other emitters along the waveguide.     (b)  Excitation transfer between two atoms as function of time as predicted from the full model (continuous lines) and the effective moving cavity model $H_{\rm mc}$ (dashed lines). In this example the emitters are detuned by $\delta/\Omega_r=-0.096$ and are separated by $|x_2-x_1|/\lambda=6$, which corresponds to a propagation time of $\tau= 754 \Omega_r^{-1}$ for an acoustic wave frequency of  $\Omega/\Omega_r=0.05$.  
   The other parameters are $V_a/E_r=0.5$, $\Delta x/\lambda=2$. The  bare coupling strength is fixed to $g_0/\Omega_r=0.007$. (c) Plot of the transfer probability $p_e^{(2)}(T_f)$ as a function of the width $\Delta x$ and the strength $V_a$ of the acoustic potential. This plot is obtained from the effective cavity model and the other parameters are the same as in (b). }
 \label{Fig5_MovingWell}
\end{figure}

In the examples discussed so far we have considered the weak-coupling regime, where the emitted photons extend over many acoustic wavelengths and a Markovian description of emitter-waveguide interactions applies. As the coupling strength increases or much shorter acoustic pulses are used this picture changes and for $g_0\sim\Omega_r$ a coherent exchange of excitations between an emitter and a photon residing inside a single potential well becomes possible. In this strong coupling regime, the nature of emitter-photon interactions and the described photon-dragging effects change completely and gives rise to new mechanisms for communicating between separated emitters.  

To illustrate this point, we focus again on the scenario, where the frequency of the emitters lies within the band gap, $\omega_{eg}<\omega_c$. 
However, instead of a continuous wave, we now consider a short acoustic pulse
\begin{equation}\label{localized_pot}
V(x,t)= -V_a  \cos[k_a(x-v t)] e^{-\frac{(x-vt)^2}{2(\Delta x)^2}},
\end{equation}
where the extent of the wave packet, $\Delta x$, is in the order of a few wavelengths. As shown in Fig.~\ref{Fig5_MovingWell}(a) for the static case $v=0$, this wave packet creates a localized potential well, which for sufficiently strong $V_a$ induces a set of spectrally isolated bound photonic states below the band edge. To model these states also at a finite speed of sound, it is convenient to change to a co-moving frame via the unitary transformation $\tilde H_w= TH_wT^\dag+i\hbar\dot T T^\dag$, where 
\begin{equation}\label{comoving frame tran}
T=e^{i\hat p vt}=e^{\hbar vt\int dx  \psi^{\dagger}(x)\frac{\partial}{\partial x} \psi(x)}.
\end{equation}
The binding energies $E_n=\hbar\omega_n$ and bound state wavefunctions $\phi_n(x)$ of the resulting time-independent Hamiltonian $\tilde H_w$ are then  solutions of the effective Schr\"odinger equation 
\begin{equation}\label{well_states}
(E_n-\hbar\omega_c) \phi_n(x)=  \left[-\frac{\hbar^2}{2m^*}\frac{\partial^2}{\partial x^2}+V(x)+i\hbar v\frac{\partial}{\partial x}\right]\phi_n(x).
\end{equation}
In the laboratory frame these photonic states are dragged along by the acoustic wave, resulting in {\it moving bound states}.
By assuming that the emitters are tuned close to the resonance of the lowest bound state with energy $E_0=\hbar \omega_0$ and wavefunction $\phi_0(x)$, we can then derive an effective moving cavity model, 
\begin{equation}\label{eq:Hcavity}
\begin{split}
H_{\rm mc}(t)= &\hbar \omega_{eg} |e\rangle\langle e|+\hbar \omega_{0}a_{0}^{\dagger}a_{0}\\&+ \hbar \sum_i \left[ g_i(t)a_{0}^{\dagger}\sigma^i_-+g^*_i(t)a_{0}\sigma^i_+\right],
\end{split}
\end{equation}
where $a_0$ and $a_0^\dag$ are photon annihilation and creation operators and  $g_i(t)=g \phi_{0}(x_i-vt)$ is the effective coupling strength between a single bound photon and the $i$-th emitter. This single-mode model is valid for  $|g_i|,|\Omega |\ll |\omega_{eg}-\omega_{n\ne 0}|,|\delta|$, i.e., as long as transitions to other bound or continuum states can be neglected. 

The Hamiltonian $H_{\rm mc}$ is formally equivalent to models that are used in the context of atomic cavity QED to describe the interaction of multiply Rydberg atoms flying through a single resonator~\cite{Scully,Brune,Raimond,Plenio,Messina}. However, here the roles are reversed, allowing the successive interaction of fixed emitters with a common cavity mode that is carried by the acoustic wave packet along the waveguide. In Fig.~\ref{Fig5_MovingWell}(b) we show how this acoustic conveyor belt can be used for implementing a state transfer protocol between two   emitters with $x_2>x_1$. In this example, the first emitter is initially prepared in the excited state $|e\rangle$ and we are interested in the excitation probability of the second emitter, $p_e^{(2)}(T_f)$, at a final time $T_f$,  once the acoustic wave has left the interaction region. The frequencies of both emitters are set to $\omega_{eg}\simeq \omega_c -0.096 \Omega_r$, which matches the frequency of the lowest photon bound state for a value of $V_a/E_r=0.5$. From the plot in Fig.~\ref{Fig5_MovingWell}(b) we see an almost perfect transfer of the excitation between the two emitters, where the delay between photon emission and reabsorption just corresponds to the propagation time $\tau=(x_2-x_1)/v$. We also find a very good agreement between the numerical simulations of the full and the effective model, as expected for the considered parameter regime, $|g_i| \ll|\omega_{eg}-\omega_{n>0}|$.

In the example above, the potential parameters $V_a$ and $\Delta x$ have been chosen to achieve perfect resonance conditions, $\omega_{eg}=\omega_0$, and to obtain a coupling $g_i(t)$ satisfying $\int_0^{T_f}g_i(t) dt=\pi$, in order to realize a complete transfer between the photon and the emitter.  In Fig.~\ref{Fig5_MovingWell}(c) we plot the same state-transfer probability for varying potential parameters. This plot demonstrates  that changing the strength and the width of the acoustic wave packet already  provides enough flexibility for finetuning the emitter-photon interaction, assuming that $\omega_{eg}$, $g$ and the speed of sound are fixed.

\section{Directional photon-emitter interactions in 2D}\label{sec:2D}

  \begin{figure}
  \centering
  \includegraphics[width=\columnwidth]{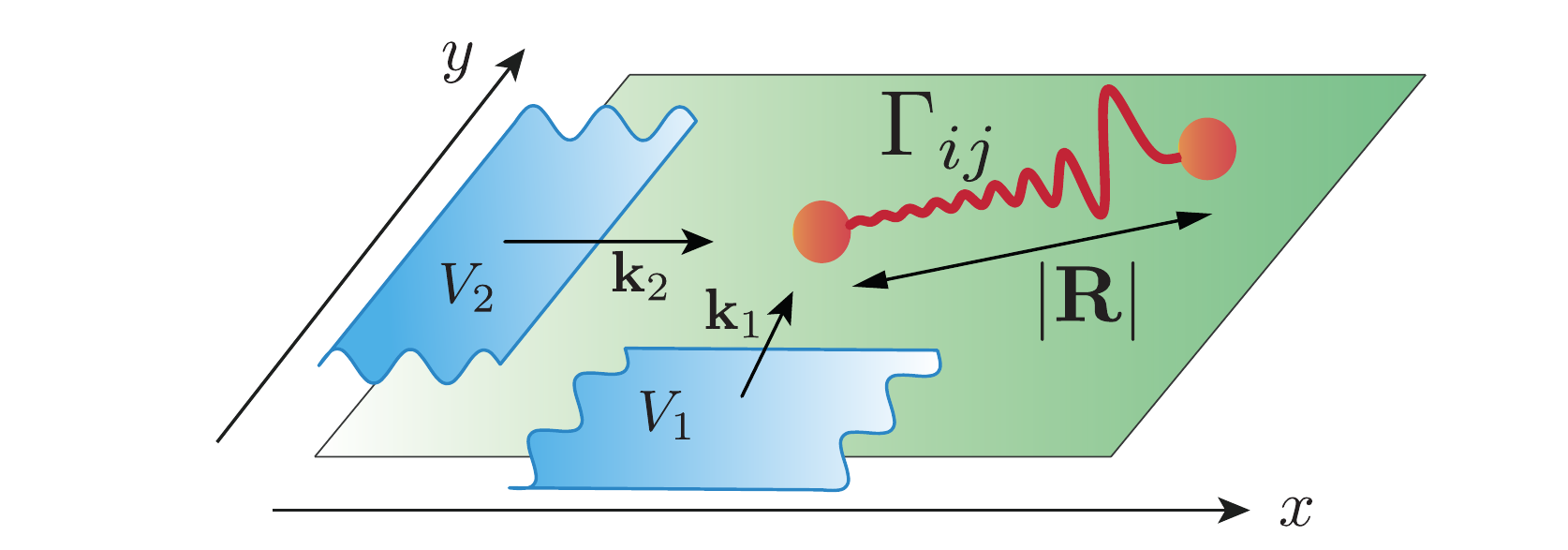}
   \caption{\textit{Acousto-optical waveguides in 2D.}---Sketch of 2D waveguide QED setting, where two emitters are coupled to photons confined along the $x-y$ plane. Acoustic waves create travelling lattice potentials with strengths $V_1$ and $V_2$ and wavevectors ${\bf k}_1$ and ${\bf k}_2$. The potential strengths and wavevectors can be adjusted to induce long-range emitter-emitter interactions along a chosen direction.}
 \label{Fig6_2DWaveguide}
\end{figure}

For the implementation of extended on-chip quantum networks it would be preferential to arrange the emitters in 2D lattices instead of along 1D arrays to achieve a higher degree of connectivity and an improved scalability.  However, photons emitted into 2D waveguides quickly spread into all directions and for two emitters separated by only several wavelengths, the ability to deterministically exchange photons becomes vanishingly small. In this section we show that the mechanism of acoustic emission control can be used to overcome this problem and to achieve fully directional emitter-emitter interactions even in a 2D scenario.

\subsection{Photon emission in 2D acousto-optical waveguides}

In the following we generalize our previous analysis to the case of a 2D optical waveguide, where the photons are strongly confined along the $z$-axis, but propagate freely in the $x-y$ plane (see Fig.~\ref{Fig6_2DWaveguide}). In this case, the Hamiltonian for the guided optical modes reads
\begin{equation}\label{eq:Hf2D}
H_w(t)=\int d^2 {\bf r}  \, \psi^\dag({\bf r})\left(\hbar \omega_c-\frac{\hbar^2}{2m^*}\nabla_{\bf r}^2+V({\bf r},t)\right)\psi({\bf r}),
\end{equation}
where ${\bf  r}=(x,y)$,   and $V({\bf r},t)$ is the potential for the photons generated by acoustic waves inside the 2D waveguide structure. In the examples below we restrict ourselves to combinations of two orthogonal plane waves,
\begin{equation}
V({\bf r},t)=V_1\cos{\left({\bf r}\cdot {\bf k}_{1} -\Omega_1 t\right)}+V_2\cos{\left({\bf r}\cdot {\bf k}_{2}-\Omega_2 t\right)},
\end{equation}
where ${\bf k}_1\perp {\bf k_2}$. 
However, all the results can be generalized to other configurations as well.    

To evaluate the emission characteristic of a single emitter under the influence of this modulation, we extend the Bloch-Floquet theory developed in  Sec.~\ref{Sec Bloch-Floquet} to two dimensions (see App.~\ref{AppB} for more details). From this analysis we obtain the quasi-energy bands $\tilde \omega_n({\bf k})$ within the first BZ defined by ${\bf k}_{1}$ and ${\bf k}_{2}$. 
Spontaneous emission occurs for all wavevectors where the resonance condition  
\begin{equation}\label{eq:Res2D}
\omega_{eg}=\tilde \omega_{n}({\bf k})+\Omega_1 \ell+\Omega_2 \ell',
\end{equation} 
is satisfied for a pair of Floquet indices $\ell$ and $\ell^\prime$. This condition defines a set of isoenergetic lines in the first BZ. For simplicity we focus in the remainder of the discussion on the regime where the acoustic potential is already sufficiently strong such that the emission is dominated by resonances in the lowest quasi-energy band ($n=1$) and with $\ell=\ell'=0$.
Under this assumption the total emission rate is given by 
 \begin{equation}\label{eq:Gamma2D}
\Gamma\simeq    \frac{g^2}{2\pi}\int_{\rm res}d \mathbf{k}\, \frac{|u_{1{\bf k}}^{(0,0)}|^2}{|\tilde {\bf v}_g(\mathbf{k})|} =  \int_0^{2\pi}  d\varphi  \, \Gamma(\varphi),
 \end{equation}
 where the $u_{n{\bf k}}^{(\ell,\ell')}$ are Bloch-Floquet expansion coefficients and $\tilde {\bf v}_g({\bf k})=\nabla_{\bf  k} \tilde \omega_{n}({\bf k})$ is the group velocity in 2D. The first integral in Eq.~\eqref{eq:Gamma2D} runs over the line of $k$-vectors satisfying the resonance condition~\eqref{eq:Res2D} for $n=1$, and $\ell=\ell'=0$. In the second expression we have introduced the polar emission rate $\Gamma(\varphi)$, which directly provides the relative fraction of photons that are emitted along the polar angle $\varphi$.

  \begin{figure}
  \centering
  \includegraphics[width=\columnwidth]{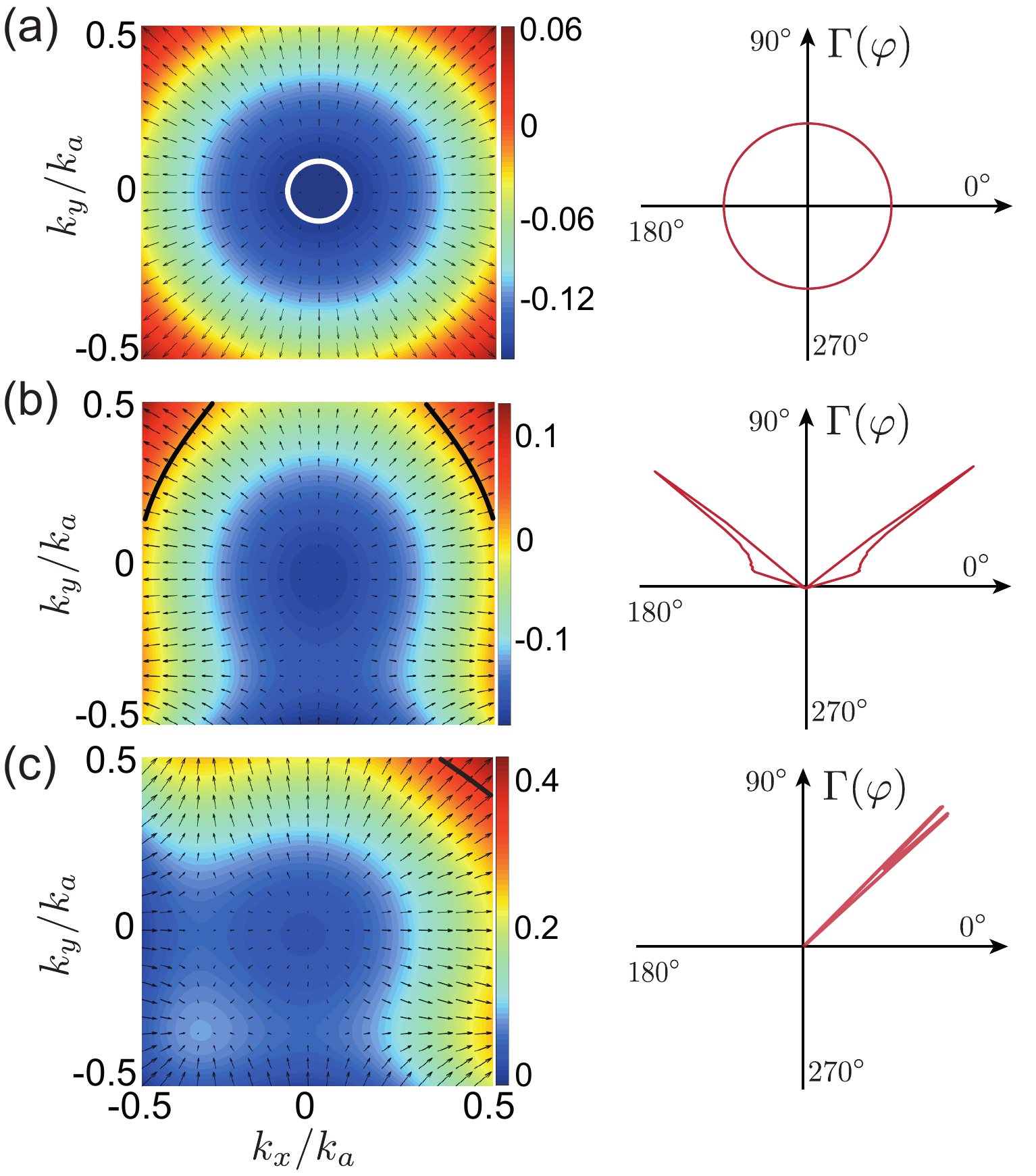}
   \caption{\textit{Directional emission in 2D.}---Plots of the lowest quasi-energy band $\tilde \omega_1({\bf k})$ of a 2D waveguide with a periodic potential with wavevectors ${\bf k}_1=(k_a,0)$ and ${\bf k}_2=(0,k_a)$. The three plots correspond to the case of (a) an unperturbed 2D waveguide ($V_{1,2}=0$), (b)  a propagating acoustic wave along $y$ ($V_{1}=0$ and $V_{2}/E_r=0.4$), and (c) two propagating waves along $x$ and $y$ ($V_{1,2}/E_r=0.4$). The black arrows indicate the direction of the group velocity $\tilde{\bf v}_g(\mathbf{k})$. The solid lines represent the isoenergetic lines defined by the resonance condition in Eq.~\eqref{eq:Res2D} for the detunings (a) $\delta/\Omega_r=0.02$, (b) $\delta/\Omega_r=0.2$ and (c) $\delta/\Omega_r=0.1$. For each of these resonance lines the plots in the right column show the resulting polar emission pattern, $\Gamma(\varphi)$.  In all plots the acoustic frequency is fixed to $\Omega_{1,2}/\Omega_r=0.2$.}
 \label{Fig7_2DBandstructure}
\end{figure}

In Fig.~\ref{Fig7_2DBandstructure} we plot the lowest quasi-energy band together with the profile of the quasi group velocity for the three basic configurations, where either no acoustic wave, a single acoustic wave or two acoustic waves propagating along orthogonal directions, are present. 
While the emission into the unperturbed waveguide is always fully isotropic~\cite{Comment}, the acoustic modulation causes again a tilting of the quasi-energy bands. As a result, the group velocities along the isoenergetic lines in the BZ are no longer equally distributed. For frequencies $\omega_{eg}$ near the upper band edge, this effect can become particularly pronounced and, as shown in the right column of Fig.~\ref{Fig7_2DBandstructure}(c), configurations can be found where the radiation pattern $\Gamma(\varphi)$ becomes highly peaked along a single direction.

\subsection{Directional emitter-emitter interactions in 2D}

For quantum networking applications as discussed in Sec.~\ref{sec_QN} above, not only a directed emission, but also the efficient reabsorption of these photons by a second emitter is important. In two or higher dimensions these two properties are not the same, since even an initially strongly focused beam can substantially spread as the distance between the emitters increases. To account for this effect, it is useful to adopt the master equation formalism developed in Sec.~\ref{sec:ME} also for 2D waveguides, where the efficiency of photon emission and reabsorption processes is directly reflected in the correlated decay rates $A_{ij}$. This allows us to quantify the suitability of a general waveguide for various quantum communication applications by a single correlation parameter
\begin{equation}
\Lambda ({\bf R})= \frac{\left|A_{12}({\bf R})\right|}{\Gamma+\Gamma_{\rm ng}} \leq 1,
\end{equation}
which takes all the relevant deviations from an ideal unidirectional waveguide (where $\Lambda=1$) into account. For instance, in the example studied in Fig.~\ref{Fig4:Entanglement}, a value of $\Lambda\approx 1$ indicates the ability to generate strong quantum correlation between two emitters, while for $\Lambda\lesssim 0.5$ this is no longer the case.

\begin{figure}
  \centering
  \includegraphics[width=\columnwidth]{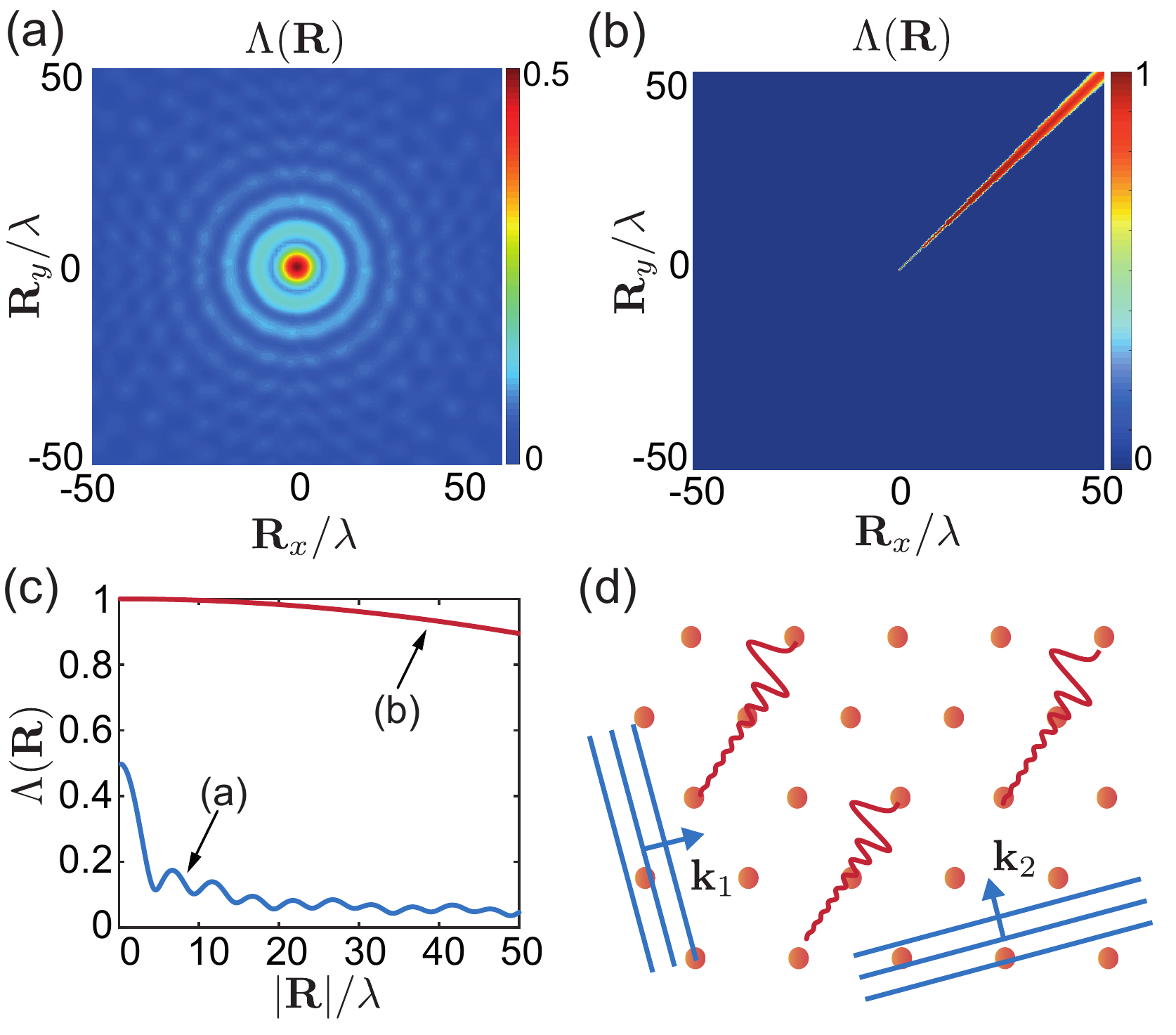}
   \caption{\textit{Directional emitter-emitter interactions in 2D.}---(a) and (b) Plot of the correlation parameter $\Lambda({\bf R})$ as a function of ${\bf R}={\bf r}_2-{\bf r}_1$ for the case of (a) an unperturbed  waveguide and (b) a waveguide modulated by two propagating acoustic waves. The detunings $\delta$ and other parameters for these two plots are the same as in the corresponding plots in Fig.~\ref{Fig7_2DBandstructure}(a) and Fig.~\ref{Fig7_2DBandstructure}(c).  For both cases the radial dependence of $\Lambda({\bf R})$ along the diagonal is plotted in (c). (d) Illustration of a triangular lattice of emitters, where a long-range unidirectional coupling along a chosen lattice direction  is established by an appropriate choice of the acoustic wavevectors ${\bf k}_1$ and ${\bf k}_2$.} 
 \label{Fig8_2DInteractions}
\end{figure}

In Fig.~\ref{Fig8_2DInteractions}(a) and (b) this correlation parameter is evaluated for a regular 2D waveguide and for a 2D waveguide in the presence of two propagating acoustic waves. As already anticipated from the corresponding plots of $\Gamma(\varphi)$ in Fig.~\ref{Fig7_2DBandstructure}, while in the static case $\Lambda({\bf R})$ is fully isotropic, in the latter case correlated emission processes are establish only along a single line, which is defined by the diagonal between the wavevectors ${\bf k}_1$ and ${\bf k}_2$. The important consequence of this directed emission is more clearly seen in Fig.~\ref{Fig8_2DInteractions}(c), which shows the radial dependence of $\Lambda({\bf R})$ along this diagonal. For a regular waveguide, we observe the typical decay, $\Lambda({\bf R})\sim 1/\sqrt{|{\bf R}|}$, as expected for photon-mediated interactions in an isotropic 2D system. Importantly, even at very small distances, the correlation parameter is always much below unity, since photons are uniformly emitted into all directions. In stark contrast, by applying acoustic control techniques strong correlations, $\Lambda({\bf R})\gtrsim 0.9$, can be established over distances that can be in the order of several tens of the acoustic wavelength, $\lambda$. 

As illustrated in Fig.~\ref{Fig8_2DInteractions}(d), this possibility to induce directional, long-range interactions even in 2D, enables the implementation of {\it fully-connected} networks of quantum nodes, where by rotating the angle of ${\bf k}_1$ and ${\bf k}_2$, emitters can interact in a unidirectional way with every other emitter in large 2D lattices.

\section{Implementation}\label{sec:Implementation}
The efficient coupling of individual emitters to propagating optical modes in nanofibers or photonic crystal waveguides has already been demonstrated~\cite{ReitzPRL2013,Hung2013,Thompson2013,Yalla2014,Arcari2014,review-lodahl,Goban2014,Hood2016,Pablo2017,Darrick_rev_mod,Lodahl_rev2} using trapped atoms as well as quantum dots. In both cases, the decay into the waveguide can substantially exceed the emission into non-guided modes, showing that the waveguide QED regime, $\Gamma_0\gg\Gamma_{\rm ng}$, can indeed be realized experimentally.
Photonic crystal waveguides can also be fabricated out of diamond~\cite{Sipahigil2016,Bhaskar2017,Burek2017,Safavi-Naeni2018}, which is particularly interesting for the present purpose. Diamond has excellent optical properties and both the speed of sound,  $v\simeq 1.7\times10^4$ m/s (for the longitudinal modes), and the refractive index, $n\simeq 2.4$, are exceptionally high compared to most other materials. At the same time there are several well-studied emitters, like  nitrogen-vacancy (NV)~\cite{Doherty2013,Childress2014} or silicon-vacancy (SiV)~\cite{Hepp2014} centers, which are ideally suited for quantum information processing applications.

Bulk or surface acoustic waves can be launched into such a photonic waveguide using either electrical interdigital transducers (IDT) or side-coupled electrodes~\cite{White1970,Sohn2018,Kinzel2015,Schulein2011,Pustiowski2015}. The resulting index modulations can be of the order of $\delta n/n_0\simeq 10^{-4}$~\cite{DeLima2005,DeLima2018,Fuhrmann2011}, which for optical frequencies of about $\omega_c/(2\pi)=400$ THz results in a potential depth of $V_a/\hbar =\omega_c \delta n/n_0\approx 2\pi \times 40$ GHz. For a diamond waveguide with an effective photon mass of $m^*\approx 1.7\times 10^{-35}$ kg and assuming an acoustic frequency of $\Omega/(2\pi)=1$ GHz ($\lambda\approx 20\,\mu$m), we obtain a recoil frequency of $\Omega_r/(2\pi) \approx 50$ GHz and $V_a/E_r\simeq 0.8$ and $\Omega/\Omega_r\simeq 0.02$. As summarized in Table~\ref{tab1}, similar values are also obtained for fused silica and silicon waveguides. 
These estimates show that already under very generic conditions, acoustic modifications of the emission characteristic, as described in this work, become experimentally accessible.

  \begin{table}
\begin{center}
\begin{tabular}{  l | c  c c c c}
   \multirow{4}{*}{Diamond} &  $\lambda[\mu m]$ & $\Omega/2\pi[{\rm GHz}]$ &$ \Omega_r/2\pi[{\rm GHz}]$ & $\Omega/\Omega_r$ &     $V_a/E_r$ \\ 
\hline
  &10 & 1.70 & 195 & 0.009 & 0.2 \\ 
   &  30 & 0.56 & 22  & 0.03 & 2\\
   &  50 & 0.34 & 8  & 0.04 & 5\\
   \hline
    \multirow{4}{*}{Silica} & 10 & 0.57 & 500 & 0.001 & 0.1 \\ 
   & 30 & 0.19 & 56  & 0.003 & 0.7\\
    & 50 & 0.11 & 20  & 0.006 & 2\\
     \hline
    \multirow{4}{*}{Silicon} & 10 & 0.84 & 74 & 0.01 & 0.5 \\ 
   & 30 & 0.28 & 8  & 0.03 & 5\\
    & 50 & 0.17 & 3  & 0.06 & 14
    \end{tabular}
    \caption{Summary of the rescaled parameters $\Omega/\Omega_r$ and $V_a/\Omega_r$ obtained for diamond, fused silica and silicon waveguides. For all examples a value of $\omega_c/(2\pi)=400$ THz and $\delta n/n=10^{-4}$ has been assumed. The other parameters used for these estimates are $n=2.4$ and $v=1.7\times 10^4$ m/s for diamond, $n=1.5$ and $v=5.7\times 10^3$ m/s for silica and $n=3.9$ $v=8.4\times 10^3$ m/s for silicon.}\label{tab1}
\end{center}
\end{table}

\subsection{Slow-light waveguides}
\begin{figure}[t]
 \centering
  \includegraphics[width=\columnwidth]{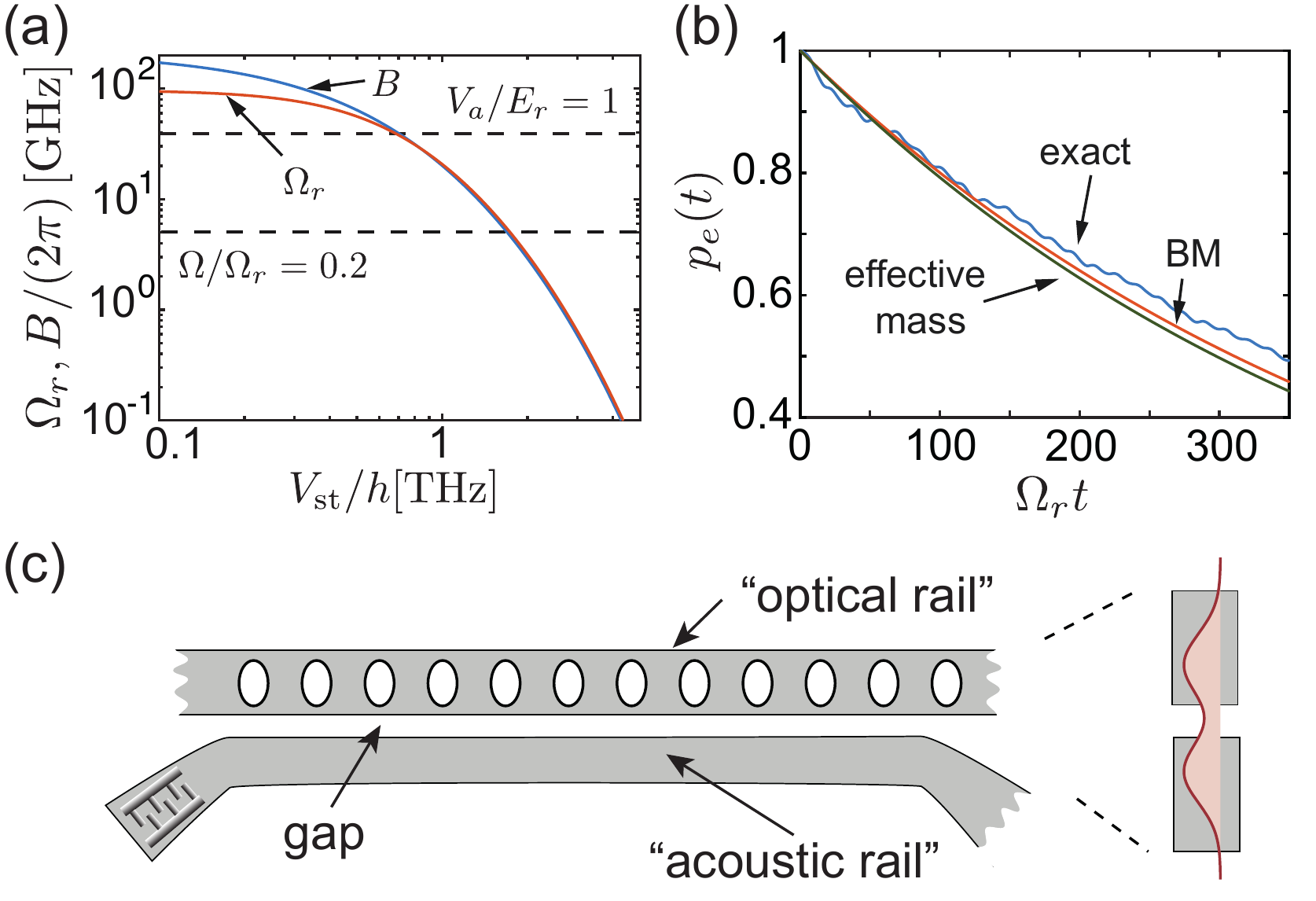}
   \caption{\textit{Implementation.}---(a) Plot of the recoil energy $\Omega_r$ and the width $B$ of the lowest band of a   static cosine potential $V(x)=V_{\rm st}\cos(2\pi x/a)$ for a silicon waveguide and $a=3\,\mu$m. The dashed lines indicate the values of the potential depth $V_a/h=40$ GHz and the acoustic frequency $\Omega/(2\pi)=1$ GHz corresponding to a wave propagating at a speed of $v\simeq 8.4\times 10^3$m/s. (b) Decay of the exited state of a single emitter in a slow-light waveguide described by the potential $V_{\rm tot}(x,t)$ in Eq.~\eqref{eq:Vtot}. Here the exact dynamics (blue line) is compared with the prediction from the Bloch-Floquet theory described in Sec.~\ref{Sec Bloch-Floquet} (green line), using the effective mass $m^*=2\hbar/(Ba^2)$ obtained from (a).  The red line is the prediction from an improved calculation detailed in App.~\ref{AppA}, using the Born-Markov (BM), but not the effective mass approximation. For this plot   $\lambda=3a=9\mu m$, $V_{\rm st}/E_r=2.5$, $g_0/\Omega_r=0.02$, $V_a/E_r=0.3$, $\Omega/\Omega_r=0.4$ and 
$\delta/\Omega_r=-0.25$, where $E_r$ and $\Omega_r$ are defined with respect to the bare effective mass, have been assumed. (c) Sketch of a dual-rail waveguide configuration, where the upper rail is used to implement the static optical crystal structure and contains the emitters, while the other, structureless rails guides the acoustic wave. By using a transverse mode profile, which extends over both waveguides, optical modes are affected by both static and acoustic potentials.}
   \label{Fig9_Implementation}
\end{figure}

To enhance acoustic effects, the group velocity of the photons in the waveguide can be further reduced by adding a static potential, $V_{\rm st}(z)$, with a periodicity $a$ that is slightly larger than the optical wavelength. For sufficiently large $V_{\rm st}$, this creates a miniband with a tight-binding dispersion relation $\omega(k)\simeq (B/2) \cos(ka)$ and an increased effective mass $m^*=2\hbar/(Ba^2)$. As illustrated in Fig.~\ref{Fig9_Implementation}(a) for the simple example $V_{\rm st}(x)=V_{\rm st}  \cos(k_{\rm st} x)$, where $k_{\rm st}=2\pi/a$, the bandwidth $B$ and therefore also the recoil frequency $\Omega_r$ can be significantly reduced compared to a regular photonic crystal waveguide.  For example, for $a=3\,\mu$m and $V_{\rm st}/h=2$ THz, which corresponds to $\delta n_{\rm st}/n_0\approx 0.005$, the resulting recoil frequency for a silicon waveguide and $\Omega/(2\pi)=1$ GHz is already reduced to $\Omega_r/(2\pi)\approx 2$ GHz.  Therefore, a potential strength of up to $V_a/E_r\simeq 20$ and ratios of $\Omega/\Omega_r\simeq 0.5$ can be reached.  At the same time, realistic coupling rates of $g_0/(2\pi)\approx 300$ MHz~\cite{Douglas2015}, are still compatible with the weak-coupling condition $g_0/\Omega_r\ll 1$, assumed in most parts of this work, and with the requirement that the resulting decay rates, $\Gamma\sim \Gamma_0$, exceed the bare decay of the emitter, $\Gamma_{\rm ng}/(2\pi)\sim 1-10$ MHz.

Note that by creating such minibands to reduce the photonic group velocity, the full optical potential entering in Hamiltonian $H_w(t)$ must be replaced by
\begin{equation}\label{eq:Vtot}
V_{\rm tot}(x,t)=V_{\rm st}(x)+ V_a \cos(k_a x-\Omega t),
\end{equation}
which in general makes the analysis of the emission processes considerably more involved (see App.~\ref{AppA}). However, as long as $a\ll\lambda$, a quasi-continuum description with an enhanced effective mass is still valid. This is demonstrated in Fig.~\ref{Fig9_Implementation}(b), where the decay of a single emitter in the presence of the combined potential given in Eq.~\eqref{eq:Vtot} is calculated for $\lambda/a=3$. Already at these ratios, the predictions from the Bloch-Floquet theory developed in Sec.~\ref{Sec Bloch-Floquet}, but assuming an enhanced effective mass, reproduce well the actual dynamics obtained from the evolution of the full model. Therefore, the implementation of minibands for optimizing the waveguide parameters can be combined with the acoustic control techniques without significantly affecting the resulting dynamics.

\subsection{Other considerations}
In our analysis we have considered so far a single waveguide that confines both the optical and the acoustic waves. This can be problematic for optimizing the photonic bandstructure, while leaving the propagation of the acoustic control signals untouched. In addition, for defects centers or quantum dots located inside the waveguide, strong strain effects give rise to large modulations of the emitter frequencies, which interfere with the described emission effects. Such and related problems can be overcome by considering dual-rail configurations~\cite{Hung2013,Chan2009}, as illustrated in Fig.~\ref{Fig9_Implementation}(c). Here one rail contains the emitters  and can be optimized to reduce the photonic group velocity, while the second rail is left unaltered to minimize the dispersion of the acoustic wave packet, reduce backscattering, etc. By using a transverse photonic mode, which has a support in both waveguides~\cite{Hung2013,Chan2009}, the photons can still be simultaneously coupled to the emitters and the acoustic waves.

\section{Conclusions}\label{sec conclusion}

In summary, we have analyzed the influence of strong acoustic waves on emitter-photon interactions in  waveguide QED. Our findings show that in particular under slow-light conditions substantial modifications of the emission dynamics and the degree of directionality can occur. These modifications can enable one to implement and control strong interactions between distant emitters.  Since these effects do not depend on specific properties of the emitter and can be tuned dynamically by simply varying the strength of the acoustic modulation, they can be applicable for a large range of quantum networking applications. Beyond the basic scenarios considered in this work, the acoustic photon dragging effects can be combined with various other lattice geometries in two or even three dimensions, where emitter-emitter interactions with different types of connectivity can be engineered. 

\emph{Note added:} At the final stage of this work, a related study about directional emission in the context of cold atoms in 2D moving optical lattices appeared~\cite{Alejandro_new}.

\section*{Acknowledgments}
The authors thank Daniele De Bernardis, Marko Loncar and Ephraim Shahmoon   for valuable discussions. This work was supported by the Austrian Science Fund (FWF) through DK CoQuS W 1210, the COST Action NQO (MP1403) and the START grant Y 591-N16, and through ONR MURI (Award No. N00014-15-1-2761), AFOSR MURI (Award  no.   FA9550-17-1-0002), CUA, NSF and Vannevar Bush Fellowship.

\appendix

\section{Bloch-Floquet theory of spontaneous emission}\label{AppA}
In this appendix we outline the derivation of the total emission rate $\Gamma$ into a 1D modulated optical waveguide, which is described by a combination of a static and a propagating potential, $V_{\rm tot}(x,t)=V_{\rm st}\cos{(k_{\rm st}x)}+ V_a \cos(k_a x-\Omega t)$. Assuming $k_a< k_{\rm st}$, the photon wavefuction can be expressed in terms of the Bloch functions $u_{nk}$ of the first BZ defined by the acoustic wavevector $k_a$. 
In the interaction picture with respect to $H_0$, the single-excitation wavefunction can then be written as $|\Psi_I(t)\rangle=\left[\tilde c_e(t)\sigma_++\sum_{nk} \tilde \phi_n(k,t) a^{\dagger }_{nk}\right] |g\rangle |{\rm vac}\rangle$ and we obtain the following set of coupled equations,
\begin{eqnarray}
&\partial_t  \tilde c_e(t) = - i   \frac{g}{\sqrt{L}}\sum_{nk} \tilde \phi_{n}(k,t)u_{nk}(0,t)e^{i\omega_{eg} t},\\
&\partial_t \tilde \phi_{n}(k,t) = - i \frac{g}{\sqrt{L}}\tilde c_e(t)u^*_{nk}(0,t)e^{-i\omega_{eg} t},
\end{eqnarray}
where $x_1=0$ has been assumed. The equations of motion of the field amplitudes can be integrated and reinserted  into the equation for $\partial_t \tilde c_e(t)$. The resulting integro-differential equation can be written as
\begin{equation}
 \begin{split}
\partial_t \tilde c_e(t) = -g^2 \int_0^t dt'   \,   G(t;t') \tilde c_e(t'),
\end{split}
 \end{equation}
 where we introduced the general correlation function for the optical field
\begin{equation}
 \begin{split}
&G(x,t; x',t')= \langle \psi(x,t) \psi^\dag(x',t')\rangle e^{i \omega_{eg} (t-t')}  \\
=& \frac{1}{L}  \sum_{nk} u_{nk}(x,t) u^*_{nk}(x',t')
   e^{i \omega_{eg} (t-t')} e^{i k (x-x')},
\end{split}
\end{equation} 
and the short notation $G(t;t')\equiv G(x=0,t;x'=0,t')$. 
 
\subsection{Continuous waveguide}\label{AppV_0}
To proceed we first consider the case $V_{\rm st}=0$, which corresponds to a continuous waveguide, as studied in Sec.~\ref{Sec. em_control}. In this limit the $u_{nk}(x,t)$ can be expanded in terms of the Bloch-Floquet  coefficients $u_{nk}^{(\ell)}$ [see Eq.~\eqref{eq:unk_decomp}], which satisfy the eigenvalue equation 
\begin{equation}\label{eig_coeff}
\sum_{\ell'}H_{\ell \ell'}u_{nk}^{(\ell')}=\tilde\omega_n(k)u_{nk}^{(\ell)},
\end{equation}
where 
\begin{equation}\label{Hproject}
H_{\ell \ell'}=\begin{cases} \Omega_r( \ell+\frac{k}{ k_a})^2-\Omega \ell , &\ell=\ell'\,\\ \frac{V_a}{2\hbar}, & |\ell-\ell'|=1,\\ 0, & \mbox{otherwise}.
\end{cases}
\end{equation}
Using this decomposition, the field correlation function can be written as
\begin{equation}\label{eq:Ftt}
 \begin{split}
G(t;t')= &\frac{1}{L}   \sum_{nk} \sum_{\ell\ell'}u_{nk}^{(\ell)} [u_{nk}^{(\ell')} ]^* e^{-i\Omega( \ell-\ell') t}  e^{i[\delta_{n}(k)-\Omega \ell'](t-t')}\\ 
\simeq &\frac{1}{L}   \sum_{nk\ell} | u_{nk}^{(\ell)} |^2  e^{i[\delta_{n}(k)-\Omega \ell](t-t')},
\end{split}
 \end{equation}
where $ \delta_{n}(k)=\omega_{eg}-\tilde\omega_n(k)$. By going from the first to the second line we have already assumed that $g_0\ll\Omega$, in which case the terms with $\ell\neq \ell'$ are fast oscillating compared to the dynamics of $\tilde c_e(t)$ and can be neglected. 

In a final step we make use of the fact that for large $|t-t'|$, i.e., on timescales relevant for the emitter dynamics,  the main contributions in Eq.~\eqref{eq:Ftt} arise from wavevectors $k_\mu$, which satisfy the resonance condition~\eqref{eq:ResonanceCondition}. By linearizing the dispersion relation in a small interval $\delta k$ around these resonances, we can approximate the correlation function by 
\begin{equation}
 \begin{split}
G(t;t')\simeq  & \sum_\mu  \frac{ | u_{n_\mu k_\mu}^{(\ell_\mu)} |^2 }{2\pi }   \int_{k_\mu-\delta k/2}^{k_\mu +\delta k/2}  dk  \, e^{-i \frac{\partial \tilde \omega_{n_\mu}}{\partial k}|_{k_\mu}  (k-k_\mu)(t-t')} \\
\simeq  & \sum_\mu  \frac{| u_{n_\mu k_\mu}^{(\ell_\mu)} |^2}{|\tilde v_{g,\mu}|}  \, \delta(t-t').
\end{split}
 \end{equation}
 This approximation corresponds to the usual Born-Markov approximation and is valid as long as $g_0$ is small compared to the width of the $n$-th quasi-energy band and away from band-edges other points where $\tilde v_g(k_\mu)\simeq  0$. 
After this simplification and by separating contributions from resonances with positive and negative group velocities, we obtain
\begin{equation}
\partial_t \tilde c_e(t) \simeq -\left(\frac{\Gamma_R}{2}+\frac{\Gamma_L}{2}\right) \tilde c_e(t),
 \end{equation}
where $\Gamma_R$ and $\Gamma_L$ are given in Eq.~\eqref{decay rate}.

\subsection{Photonic crystal waveguide}
We now consider the more general case of a photonic crystal waveguide, where $V_{\rm st}\neq 0$. For simplicity we consider here only the case where the acoustic wavelength is an interger multiple of the period of the static potential, i.e., $\lambda/a= k_{\rm st}/k_a= M\in {\mathbbm N}$. Under this assumption the original band structure generated by $V_{\rm st}$ is divided into $M$ sub-bands by the acoustic wave. Then, the Bloch functions $u_{nk}(x,t)$ can be expanded as
\begin{equation}\label{ansatzsl}
u_{nk}(x,t)=\sum_{\ell, \nu }u_{nk}^{(\ell,\nu)}e^{i(k+ k_a \nu)x}e^{-i(\tilde \omega_n(k)+\Omega \ell) t},
\end{equation}
where now an additional index $\nu$ for the Fourier expansion in $x$ must be introduced. By inserting this ansatz into Eq.~\eqref{eq:unk_dot} we obtain an eigenvalue equation for the quasi-energies $\tilde \omega_n(k)$, similar to Eq.~\eqref{eig_coeff}. The matrix elements for the corresponding Floquet Hamiltonian are
\begin{equation}\label{H0_floq}
H_{\ell\ell',\nu \nu'}=\begin{cases}  \Omega_r( \nu+\frac{k}{ k_a})^2-\Omega \ell , &\ell=\ell', \nu=\nu', \\ 
\frac{V_{\rm st}}{2\hbar},  & \ell=\ell', |\nu- \nu'|=M,\\
\frac{V_a}{2\hbar},  & \ell- \ell'= \nu- \nu' =  \pm1, \\\
 0, & \mbox{otherwise}.
\end{cases}
\end{equation}
For the derivation of the total decay rate we can then proceed as above and under the validity of the Born-Markov approximation we obtain 
\begin{equation}\label{decay rate_superlatt}
\Gamma=\sum_{\mu}\frac{| g_\mu(x_1)|^2} {|\tilde v_{g,\mu}|}.
\end{equation} 
Note that in contrast to the homogeneous waveguide the couplings $g_{\mu}(x)=g \sum_\nu u_{n_{\mu}k_{\mu}}^{(\ell_{\mu},\nu)} e^{ik_a \nu x}$  depend explicitly on the location of the emitter within the static lattice potential, $V_{\rm st}(x)$.

\section{Derivation of the master equation}\label{AppB}
For the derivation of the master equation~\eqref{eq:MasterEq} we consider the general setting of $N$ emitters located at positions $x_i$ along a 1D waveguide. In the interaction picture with respect to $H_0$, the emitter-field coupling reads
\begin{equation}
H_I(t)= \hbar g\sum_{i=1}^{N}\left[\psi^\dag(x_i,t)\sigma_-^i e^{-i\omega_{eg} t} +\sigma_+^i \psi(x_i,t) e^{-i\omega_{eg} t}\right].
\end{equation}
Under the validity of the Born-Markov approximation, we can follow the usual approach and derive an effective, time-local master equation for the  reduced density operator of the emitters, $\rho$ (see, for example, Ref.~\cite{Petruccione}). The result is of the general form 
\begin{equation}
\begin{split}
& \dot \rho =- \frac{1}{\hbar^2} \int_{0}^\infty d\tau \, {\rm Tr}_f \{ [H_I(t), [ H_I(t-\tau), \rho(t)\otimes \rho_f ]]\},
\end{split} 
\end{equation}
where $\rho_f=|{\rm vac}\rangle\langle {\rm vac}|$ and the trace is over the field degrees of freedom.  By evaluating all the individual terms, the result can be written in a compact notation as 
\begin{equation}\label{eq:appMA}
\dot \rho(t) = \sum_{i,j=1}^N A_{ij}  \left(\sigma_-^i\rho \sigma_+^j- \sigma_+^j\sigma_-^i\rho\right) +\rm H.c.,
\end{equation}
where
\begin{equation}\label{eq:AijApp}
A_{ij}= g^2 \int_0^\infty d\tau \,  G(x_j, \tau; x_i, 0).
\end{equation}
For weak enough driving strength, $\Omega_L$, the Hamiltonian for the external laser fields can be added to Eq.~\eqref{eq:appMA} without affecting the validity of this result.
Then, after identifying $\Gamma=2 {\rm Re} \{ A_{ii}\}$ and neglecting small frequency shifts $\sim {\rm Im}\{A_{ii}\}$ we recover Eq.~\eqref{eq:MasterEq}.

\subsection{Correlated decay rates in 1D and 2D}
For the evaluation of the correlated decay rates $A_{ij}$ in a 1D waveguide we can use the same set of approximations as in App.~\ref{AppA}. As a main difference, the $A_{ij}$ depend on the distance $ r_{ij}=x_j-x_i$, since
\begin{equation}
 \begin{split}
&G(x_i, \tau; x_j, 0)
\simeq \frac{1}{L}   \sum_{nk\ell} | u_{nk}^{(\ell)} |^2  e^{i[\delta_{n}(k)-\Omega \ell]\tau}e^{i[k+k_a \ell]r_{ij}}\\
&\simeq  \sum_\mu  \frac{| u_{n_\mu k_\mu}^{(\ell_\mu)} |^2}{|\tilde v_{g,\mu}|} e^{i[k_\mu+k_a \ell_\mu]r_{ij}} \, \delta\left(\tau - \frac{r_{ij}}{\tilde v_{g,\mu} }\right).
\end{split}
 \end{equation}
Therefore, non-vanishing contributions to $A_{ij}$ arise only from resonances where $r_{ij}$ and $\tilde v_{g,\mu}$ are co-aligned. In Eq.~\eqref{eq:Aij}, this fact is accounted for by the step-function $\theta[ \tilde v_{g,\mu}r_{ij}]$.

The derivation of the master equation can be readily generalized to 2D waveguides, by using in Eq.~\eqref{eq:AijApp} the corresponding 2D correlation function $G({\bf r}_i, \tau; {\bf r}_j, 0)$ for the evaluation of the $A_{ij}$. In this case and defining $\mathbf{R}_{ij}=\mathbf{r}_j-\mathbf{r}_i$, the general expression for the correlation decay rates reads
 \begin{equation}
A_{ij}=\frac{g^2}{2\pi} \sum_{n\ell\ell'} \int_{\rm res}d \mathbf{k}\frac{|u_{n{ \bf k}}^{(\ell,\ell')}|^2e^{i\mathbf{k}\cdot \mathbf{R}_{ij}}}{|\mathbf {\tilde v}_g(\mathbf{k})|}\theta[ \mathbf {\tilde v}_g(\mathbf{k})\cdot\mathbf{R}_{ij}],
 \end{equation}
where for each set of indices $n$, $\ell$ and $\ell'$ the ${\bf k}$-integration runs over the resonance lines defined by Eq.~\eqref{eq:Res2D}. Note that for an isotropic waveguide and without the acoustic modulation,
\begin{equation}
A_{ij} = \frac{g^2 k_r}{2v_g(k_r)}\left[ J_0(k_r |{\bf R}_{ij}|)+ i H_0(k_r |{\bf R}_{ij}|) \right],
\end{equation}
where $J_0(x)$ and $H_0(x)$ denote the zeroth order Bessel and Struve functions, respectively. From the asymptotic expansion of these functions one obtains the characteristic decay, $A_{ij}\sim 1/\sqrt{|{\bf R}_{ij}|}$, for photon-mediated interactions in 2D.



\begin{thebibliography}{99}
\bibitem{Comment} Note that in also in regular 2D photonic crystal structures the emission can become highly peaked~\cite{Alejandro_2d_prl,Alejandro_2d_pra}, but since the symmetry is not broken, this emission still occurs along four diagonals.
\end{thebibliography}
\end{document}